\documentclass[aps,pre,twocolumn,showpacs,superscriptaddress]{revtex4}

\usepackage{natbib}
\usepackage{longtable}
\usepackage{enumerate}
\usepackage{amsmath,amssymb,bbm}
\usepackage{color}

\usepackage{graphicx}

\newcommand{\fracd}[2]{
\displaystyle{
\frac{ \displaystyle{#1} }{ \displaystyle{#2} }
}
}

\newcommand{\diff}[1]{\mathrm{d}{#1}}

\newcommand{\taub}{\tau_{\mathrm{bulk}}}
\newcommand{\htau}{\hat{\tau}}
\newcommand{\htaub}{\hat{\tau}_{\mathrm{bulk}}}

\newcommand{\ii}{{\dot{\imath}}}
\newcommand{\jj}{{\dot{\jmath}}}

\renewcommand{\c}{\mathbf{c}}
\newcommand{\ci}{\c_{\ii}}

\newcommand{\vbf}[1]{\mathbf{#1}}
\newcommand{\dt}{\Delta t} 
 \newcommand{\tend}{t_{\mathrm{end}}}
\newcommand{\dx}{\Delta x}
\newcommand{\rhor}{\rho^{\circ}}

\newcommand{\vvel}{\vbf{v}}
\newcommand{\vveleq}{\vbf{v}^{*}}

\newcommand{\vk}{v_{\ak}}

\newcommand{\x}{\vbf{x}}

\renewcommand{\xi}{x_{\ai}}
\newcommand{\xj}{x_{\aj}}

\newcommand{\tx}{\vbf{g}}

\newcommand{\ele}{\ell} 
\newcommand{\bfl}{\boldsymbol{\ele}}
\newcommand{\li}{\ele_{\ai}}
\newcommand{\lj}{\ele_{\aj}}
\newcommand{\lk}{\ele_{\ak}}

\newcommand{\Li}{L_{\ai}}
\newcommand{\Lj}{L_{\aj}}
\newcommand{\Lk}{L_{\ak}}

\newcommand{\ai}{1}
\newcommand{\aj}{2}
\newcommand{\ak}{3}

\newcommand{\e}{\mathbf{e}}
\newcommand{\ei}{\e_{\ai}}
\newcommand{\ej}{\e_{\aj}}
\newcommand{\ek}{\e_{\ak}}

\newcommand{\fzv}{\vbf{b}}
\newcommand{\fz}{b}
\newcommand{\hfz}{\hat{b}}

\newcommand{\bfM}{\vbf{M}}
\newcommand{\bfS}{\vbf{S}}
\newcommand{\bfhS}{\vbf{\check{S}}}
\newcommand{\hS}{\check{S}}

\newcommand{\f}{f}
\newcommand{\bff}{\vbf{\f}}
\newcommand{\bffq}{{\bff}^{\mathrm{eq}}}
\newcommand{\fii}{\f_{\ii}}
\newcommand{\fq}{\f^{\mathrm{eq}}}

\newcommand{\fqi}{\fq_{\ii}}

\newcommand{\fj}{\f_{\jj}}
\newcommand{\fqj}{\fq_{\jj}}

\newcommand{\m}{m}
\newcommand{\bfm}{\vbf{\m}}
\newcommand{\mq}{\m^{\mathrm{eq}}}
\newcommand{\bfmq}{{\bfm}^{\mathrm{eq}}}

\newcommand{\bfOmega}{\boldsymbol{\Omega}}
\newcommand{\wwi}[1]{\omega^{\text{#1}}_{\ci}}
\newcommand{\cs}{{c_{\mathrm{s}}}}

\newcommand{\sbgk}{\text{\tiny{\BGK}}}
\newcommand{\smrt}{\text{\tiny{\MRT}}}
\newcommand{\sLB}{\text{\tiny{\LB}}}

\newcommand{\thry}{\text{\tiny{TH}}}

\newcommand{\crcl}{\text{\tiny{$\bigcirc$}}}

\newcommand{\dpl}{{(\nabla p)}_{3}}

\newcommand{\lr}{\ensuremath{\text{l.r.}}}
\newcommand{\hr}{\ensuremath{\text{h.r.}}}
\newcommand{\extrap}{\ensuremath{\text{extrap}}}

\newcommand{\LB}{\ensuremath{\text{LB}}}
\newcommand{\BGK}{\ensuremath{\text{BGK}}}
\newcommand{\MRT}{\ensuremath{\text{MRT}}}
\newcommand{\LBBGK}{\ensuremath{\text{LB-BGK}}}
\newcommand{\LBMRT}{\ensuremath{\text{LB-MRT}}}
\newcommand{\TRT}{\ensuremath{\text{TRT}}}

\newcommand{\muCT}{\ensuremath{\text{$\mu$-CT}}}
\newcommand{\BC}{\ensuremath{\text{BC}}}
\newcommand{\PBC}{\ensuremath{\text{PBC}}}
\newcommand{\SBB}{\ensuremath{\text{SBB}}}

\newcommand{\mD}{\mathrm{mD}}

\renewcommand{\Re}{\ensuremath{\mathrm{Re}}}
\newcommand{\Ma}{\ensuremath{\mathrm{Ma}}}

\newcommand{\Lin}{L_{\mc{I}}}
\newcommand{\Lout}{L_{\mc{O}}}
\newcommand{\Ls}{L_{\mc{S}}}

\newcommand{\Cinp}{\mc{C}(\Lin\!+\!1)}
\newcommand{\Cinm}{\mc{C}(\Lin)}
\newcommand{\Coutp}{\mc{C}(\Lk\!-\!\Lout+1)}
\newcommand{\Coutm}{\mc{C}(\Lk\!-\!\Lout)}

\newcommand{\mc}[1]{\mathcal{#1}}

\newcommand{\pwr}[1]{\!\times\!10\sp{#1}}
\newcommand{\pwrr}[1]{10\sp{#1}}

\newcommand{\mtA}{\ensuremath{\text{\sc method~A}}}
\newcommand{\mtB}{\ensuremath{\text{\sc method~B}}}
\newcommand{\mtC}{\ensuremath{\text{\sc method~C}}}
\newcommand{\mtD}{\ensuremath{\text{\sc method~D}}}

\begin{document}

\title{Quantitative analysis of numerical estimates for the permeability of porous media from 
lattice-Boltzmann simulations}

\author{Ariel Narv\'{a}ez} \affiliation{ Department of Applied
  Physics, TU Eindhoven, P.O. Box 513, NL-5600MB Eindhoven, The
  Netherlands } \affiliation{ Institute for Computational Physics,
  University of Stuttgart, Pfaffenwaldring 27, D-70569 Stuttgart,
  Germany }

\author{Thomas Zauner} \affiliation{ Institute for Computational
  Physics, University of Stuttgart, Pfaffenwaldring 27, D-70569
  Stuttgart, Germany }

\author{Frank Raischel} \affiliation{ Institute for Computational
  Physics, University of Stuttgart, Pfaffenwaldring 27, D-70569
  Stuttgart, Germany }

\author{Rudolf Hilfer} \affiliation{ Institute for Computational
  Physics, University of Stuttgart, Pfaffenwaldring 27, D-70569
  Stuttgart, Germany } \affiliation{ Institute for Physics, University
  of Mainz, D-55099 Mainz, Germany }

\author{Jens Harting} \affiliation{ Department of Applied Physics, TU
  Eindhoven, P.O. Box 513, NL-5600MB Eindhoven, The Netherlands }
\affiliation{ Institute for Computational Physics, University of
  Stuttgart, Pfaffenwaldring 27, D-70569 Stuttgart, Germany }

\date{\today}


\begin{abstract}
  During the last decade, lattice-Boltzmann (\LB) simulations have
  been improved to become an efficient tool for determining the
  permeability of porous media samples.  However, well known
  improvements of the original algorithm are often not
  implemented. These include for example multirelaxation time schemes
  or improved boundary conditions, as well as different possibilities
  to impose a pressure gradient. This paper shows that a significant
  difference of the calculated permeabilities can be found unless one
  uses a carefully selected setup.
  We present a detailed discussion of possible simulation setups and
  quantitative studies of the influence of simulation parameters.
  We illustrate our results by applying the algorithm to a Fontainebleau
  sandstone and by comparing our benchmark studies to other numerical
  permeability measurements in the literature.
\end{abstract}

\pacs{
47.11.-j     
91.60.Np     
47.56.+r     
}

\maketitle

\section{Introduction}
The accurate numerical simulation of fluid flow in porous media is
important in many applications ranging from hydrocarbon production and
groundwater flow to catalysis and the gas diffusion layers in fuel
cells~\cite{Hil96}. Examples include the behavior of liquid oil and
gas in porous rock~\cite{PhysRevB.45.7115}, permeation of liquid in fibrous
sheets such as paper~\cite{koponen:3319}, determining flow in
underground reservoirs and the propagation of chemical contaminants in
the vadose zone~\cite{2005EG:DN,citeulike:481254}, assessing the
effectiveness of leaching processes~\cite{2007SUM:AW} and optimizing
filtration and sedimentation operations~\cite{2003IJMP:GCB}. An
important and experimentally determinable property of porous media is
the permeability, which is highly sensitive to the underlying
microstructure. Comparison of experimental data to numerically
obtained permeabilities can improve the understanding of the influence
of different microstructures and assist in the characterization of the
material.

Before the 1990's the computational power available was very
limited restricting all simulations either to small length scales or
low resolution of the microstructure. Shortly after its introduction
lattice-Boltzmann (\LB) simulations became
popular~\cite{bib:ferreol-rothman,1990PhFl.2.2085C,bib:qian-dhumieres-lallemand}
as an alternative to a direct numerical solution of the Stokes
equation~\cite{PhysRevB.46.6080,MAKHT02} for simulating fluid flow in complex
geometries. 
Historically, the \LB\ method was developed from the lattice gas
automata~\cite{bib:qian-dhumieres-lallemand,bib:succi-01}. In contrast
to its predecessor, in the \LB\ method the number of particles in each
lattice direction is replaced with the ensemble average of the single
particle distribution function, and the discrete collision rule is
replaced by a linear collision operator.

In the \LB\ method all computations involve local variables so that it can
be parallelized easily~\cite{MAKHT02}.  With the advent of more powerful computers it
became possible to perform detailed simulations of flow in artificially
generated geometries~\cite{koponen:3319}, tomographic reconstructions of
sandstone samples~\cite{bib:ferreol-rothman,Martys99largescale,
MAKHT02,bib:jens-venturoli-coveney:2004,Ahren06}, or fibrous sheets of
paper~\cite{koponen:716}.

The accuracy of \LB\ simulations of flow in porous media depends on
several conditions. These include the resolution of the discretization of the
porous medium, proper boundary conditions to drive the flow and to implement the
solid structure or the choice of the collision kernel. Even though advanced
boundary conditions, discretization methods, as well as higher order \LB\ 
kernels have been developed and are common in the literature, it is
surprising to the authors that they only found limited applications so far. In
particular for commercial applications a three-dimensional implementation with
19 discrete velocities and a single relaxation time linearized collision
operator is still the de-facto standard to calculate stationary velocity fields
and absolute permeabilities for porous media~\cite{biswal:061303}. Here, the
flow is usually driven by a uniform body force to implement a pressure gradient
and solid surfaces are generated by simple bounce back boundary conditions.

The present work is motivated by the question whether permeabilities
calculated by this standard \LB\ approach can be considered to be
accurate. In particular, it is important to understand where the
limits of this method are and how the accuracy can be increased. We
quantify the impact of details of the implementation by studying 3D
Poiseuille flow in pipes of different shape and resolution and
comparing the simulation results to analytical solutions. This allows
to demonstrate how simple improvements of the simulation paradigm can
lead to a substantial reduction of the error in the measured
permeabilities. These include a suitable choice of the relaxation
parameter $\tau$ and the application of the multirelaxation time
method in order to ascertain a minimal unphysical influence of the
fluid viscosity on the permeability. Further, a correct implementation
of the body force to drive the flow together with suitable in- and
outflow boundaries is mandatory to avoid artifacts in the steady state
velocity field. Finally, the small compressibility of the \LB\ fluid
requires a proper determination of the pressure gradient in the
system. If these details are taken care of, it is shown that the
\LB\ method is well suitable for accurate permeability calculations of
stochastic porous media by applying it to discretized micro
computer-tomography (\muCT) data of a Fontainebleau sandstone.

\section{Simulation method}
\label{simmeth:sec}
The Boltzmann equation
\begin{equation}
\label{boltz:eq}
\frac{\partial}{\partial t } \f(\x,\c ,t) + 
\c \cdot  \nabla \f(\x,\c ,t)  = \Omega(\f(\x,\c ,t))
\end{equation}
describes the evolution of the single particle probability density
$\f(\x,\vbf{c},t)$, where $\x\in\mathbbm{R}^{3}$ is the position
vector, $\c\in\mathbbm{R}^{3}$ is the velocity vector, $t \in
\mathbbm{R}$ is the time, and $\Omega(\f(\x,\c,t))$ is the collision
operator. While discretizations on unstructured grids 
exists~\cite{PhysRevE.58.R4124,PhysRevE.68.016701}, 
they are not widely used and typically the position $\x$ is discretized on a
structured cubic lattice, with lattice constant $\dx$. 
The time is discretized
using a time step $\dt$ and the velocities are discretized into a
finite set of vectors $\ci$ with $i=1, \ldots, N$, called lattice
velocities, where the finite integer $N$ varies between
implementations. In this work we exclusively use the so-called D3Q19
lattice, where $N=19$ velocities are used in a three dimensional
domain~\cite{bib:qian}.
A cubic lattice with basis $\e_k\in \mathbbm{R}^3$, $k=1,2,3$ is embedded into
$\mathbbm{R}^3$ using the coordinate function $\tx:\mathbbm{N}^3 \mapsto
\mathbbm{R}^3$ to map the lattice nodes $\bfl \in  \mathbbm{N}^3$ to position
vectors $\tx(\bfl) \in  \mathbbm{R}^3$. The computational domain is a
rectangular parallelepiped denoted as
\begin{equation}
\label{ldomain}
\mc{L}=\{ \bfl \in \mathbbm{N}^3 : 
1 \leq \ele_k \leq L_{k} ; \, k=1,2,3 \},
\end{equation}
where $L_k\in \mathbbm{N}^3$ are its dimensionless side-lengths. 
See Fig.~\ref{fig:permcalc} for a visualization.
Physical quantities $w$ such as pressure or density on the lattice are
abbreviated as $w(\bfl)=w(\tx(\bfl))$.
We introduce the vector notation 
$\bff (\bfl,t) =\left(\f_{1}(\bfl,t) ,\ldots , \f_{N}(\bfl,t)\right)$, where
the components are the probabilities calculated as
\begin{equation}
\label{sdistri}
\fii(\bfl,t)=
\int_{\mathbbm{W}(\bfl)}\int_{\mathbbm{B}(\ii)}\f(\x,\c,t)\,\diff{\c}\,\diff{\x}.
\end{equation}
Here, $\mathbbm{W}(\bfl)\subset\mathbbm{R}^{3}$ is the finite volume
associated with the point $\tx(\bfl)$ and
$\mathbbm{B}(\ii)\subset\mathbbm{R}^{3}$ is the volume in velocity space
given by lattice velocity $\ci$. 
The macroscopic density $\rho(\bfl,t)$ and 
velocity $\vvel(\bfl,t)$ are obtained from $\fii(\bfl,t)$ as
\begin{eqnarray}
\label{rho_lB}
\rho(\bfl,t)&=&\rhor \sum_{\ii=1}^{N}\fii(\bfl,t),\\
\label{mom_lB}
\vvel(\bfl,t)&=& 
 {\sum_{\ii=1}^{N}\fii(\bfl,t) \, \ci }\Big/{ \sum_{\ii=1}^{N}\fii(\bfl,t)  },
\end{eqnarray}
where $\rhor$ is a reference density. The pressure is given by
\begin{equation}
\label{p_lB}
p(\bfl,t) = \cs^2 \, \rho(\bfl,t), 
\end{equation}
with the speed of sound~\cite{bib:qian-dhumieres-lallemand,bib:succi-01} 
\begin{equation}
\label{cs}
\cs = \frac{1}{\sqrt{3}}\left(\frac{\dx}{\dt}\right).
\end{equation}

Discretization of Eq.~\eqref{boltz:eq} provides the basic system of difference 
equations in the \LB\ method 
\begin{equation}
\label{discboltz}
\fii(\bfl+\Delta\bfl_{\ii},t+\dt)-\fii(\bfl,t)=\dt\,\Omega_{\ii}(\bfl,t),
\end{equation}
with 
$\Delta\bfl_{\ii}=\ci \dt /\dx$ and the initial condition $\fii(\bfl,0)=1/N$ (for $t=0$). 
The generally nonlinear collision operator is approximated using the linearization
\begin{equation}
\label{coll:lin}
\Omega_{\ii}(\bfl,t)=\sum_{\jj=1}^{N}S_{\ii\jj}(\fj(\bfl,t)-\fqj(\bfl,t)), 
\end{equation}
around a local equilibrium probability function
\mbox{$\bffq(\bfl,t)=(\fq_{1}(\bfl,t), \ldots ,\fq_{N}(\bfl,t))$},
with a $N \times N$ collision matrix
$\bfS$~\cite{bib:higuera-succi-benzi,Ladd01}.

The simplest approach to define the collision matrix uses a single
relaxation time with time constant $\tau$,
\begin{equation}
\label{sbgk}
S_{i\,j}=-\fracd{1}{\tau}\,\delta_{i\,j}, 
\end{equation}
where $\delta_{i\,j}$ is the Kronecker delta. This single relaxation
time (\LBBGK) scheme is named after the original work of
Bhatnagar, Gross and
Krook~\cite{bib:chen-chen-martinez-matthaeus,bib:bgk}. Within the
\LBBGK\ method, $\bffq(\bfl,t)$ is approximated by a second order
Taylor expansion of the Maxwell distribution~\cite{bib:DHae:04},
\begin{equation}
\label{qian}
\fqi(\bfl,t)\!=\!
\frac{\rho\wwi{}}{\rhor}\!
\left( 1 \!+\!
\frac{\vveleq\cdot\ci}{\cs^{2}} \!+\!
\frac{(\vveleq\cdot\ci)^{2}} {2\cs^{4}}\!-\!\frac{\vveleq\cdot\vveleq}{2\cs^{2}} \right)\!.
\end{equation}
If
external forces are absent, the equilibrium velocity 
is defined as
$\vveleq(\bfl,t)=\vvel(\bfl,t)$ from Eq.~\eqref{mom_lB}. As
explained further below, $\vveleq(\bfl,t)$ and $\vvel(\bfl,t)$ may
differ from Eq.~\eqref{mom_lB} if an external acceleration is present.
The numbers $\wwi{}$ are called lattice weights and differ with
lattice type, number of space dimensions and number of discrete
velocities $N$. See~\cite{bib:qian-dhumieres-lallemand} for a
comprehensive overview on different lattices.

An alternative approach to specify the collision matrix is the multirelaxation time
(\MRT) method. Here, a linear transformation $\bfM$ is chosen
such that the moments
\begin{equation}
m_{\ii}(\bfl,t)=\sum_{\jj}^{N}{M_{\ii\,\jj}\, \f_{\jj}(\bfl,t)}
\end{equation}
represent hydrodynamic modes of the problem. We use the definitions
given in~\cite{2002RSPTA.360..437D}, 
where $\m_{1}(\bfl,t)$ is the density defined in Eq.~\eqref{rho_lB},
$\m_{2}(\bfl,t)$ represents the energy, $\m_{\ii}(\bfl,t)$ with
$i=4,6,8$ the momentum flux and $\m_{\ii}(\bfl,t)$, with
$i=10,12,14,15,16$ are components of the symmetric traceless stress
tensor. Introducing the moment vector
$\bfm(\bfl,t)=(\m_{1}(\bfl,t),\dots,\m_{N}(\bfl,t))$, $\bfOmega(\bfl,t)=
(\Omega_{1}(\bfl,t),\ldots ,\Omega_{N}(\bfl,t) \,)$, a diagonal matrix
$\hS_{\ii\,\jj} = \check{s}_{\ii} \,\delta_{\ii\,\jj}$, and the
equilibrium moment vector
$\bfmq(\bfl,t)=(\mq_{1}(\bfl,t),\dots,\mq_N(\bfl,t))$, we obtain
\begin{equation}
\label{mtransn}
\bfOmega(\bfl,t)=-{\bfM}^{-1}\cdot\bfhS\cdot\left(\bfm(\bfl,t)-
\bfmq(\bfl,t)\right).
\end{equation}
During the collision step the density and the momentum flux are
conserved so that $\mq_{1}(\bfl,t)=\m_{1}(\bfl,t)$ and
$\m_{\ii}(\bfl,t)=\mq_{i}(\bfl,t)$ with $i=2,4,6$. The non-conserved
equilibrium moments $\mq_{\ii}(\bfl,t)$, $i\neq 1,2,4,6$, are assumed
to be functions of these conserved moments and explicitly given e.g.
in~\cite{2002RSPTA.360..437D}. 
The diagonal element $\tau_{\ii}=1/\check{s}_{\ii}$ in the collision
matrix is the relaxation time moment $\m_{\ii}(\bfl,t)$. One has
$\check{s}_{1}=\check{s}_{4}=\check{s}_{6}=\check{s}_{8}=0$, because the
corresponding moments are conserved, $\check{s}_{2}=1/\taub$ describes
the relaxation of the energy and
$\check{s}_{10}=\check{s}_{12}=\check{s}_{14}=\check{s}_{15}=\check{s}_{16}=1/
\tau$ the relaxation of the stress tensor components.  The remaining
diagonal elements of $\bfhS$ are chosen as 
\begin{multline}
\bfhS=\mathrm{diag}(0,1/\taub,1.4,0,1.2,0,1.2,0,1.2,1/\tau,\\
1.4,1/\tau,1.4,1/\tau,1/\tau,1/\tau,1.98,1.98,1.98),
\end{multline}
to optimize the algorithm
performance~\cite{2000PhRvE..61.6546L,2002RSPTA.360..437D}. Because
two parameters $\tau$ and $\taub$ remain free, the
multirelaxation time method reduces to a ``two relaxation time''
(\TRT) method. An alternative \TRT\ implementation can be found
in~\cite{el00178,el00173}.

To apply the \LB\ method to viscous flow in porous media it is necessary to
establish its relations with hydrodynamics.
The Chapman-Enskog procedure shows that density, velocity and pressure
fulfill the Navier-Stokes equations without external forces, with a
kinematic viscosity~\cite{bib:chapman-cowling, Wolf05,bib:DHae:04,
  bib:cf.CPaLLuCMi.2006,2003PhRvE..68c6706L}
\begin{equation}
\label{visc_BGK}
\nu(\tau,\dt)=\cs^2 \dt\left(\frac{\tau}{\dt}-\frac{1}{2}\right).
\end{equation}
Combining Eq.~\eqref{visc_BGK} and Eq.~\eqref{cs} gives
\begin{equation}
\frac{\tau}{\dt}
=\frac{1}{2}+\frac{\sqrt{3} \,  \nu  }{\cs \,   \dx}
=\frac{1}{2}+ 3\frac{\, \nu \, \dt }{(\dx)^2}.
\end{equation}
Because $\nu \geq 0$, $\dx>0$, and $\dt>0$, 
it follows that $\tau/\dt \geq 1/2$.

A typical value for the pore diameter in sandstone is
$ a\approx\pwrr{-5}\,\mathrm{m}$, and for water the kinematic
viscosity and speed of sound are $\nu\approx\pwrr{-6} \,
\mathrm{m}^2 \,\mathrm{s}^{-1}$ and $\cs\approx\pwrr{3} \,
\mathrm{m} \, \mathrm{s}^{-1}$, respectively. With typical
velocities of order $v\approx\pwrr{-4}\,\mathrm{m}\,\mathrm{s}^{-1}$
the Reynolds number is $\Re = v \, a / \nu \approx
\pwrr{-3}$. Discretizing with $\dx=\pwrr{-6} \, \mathrm{m}$ gives then
$\tau/\dt =0.5017$. Because for $\tau/\dt \approx 1/2$ the \LB\ method
is known to be unstable, a direct simulation of water flow in porous
media with these parameters is not feasible.  To overcome this
impasse, one might impose $\tau/\dt=1$ and simultaneously fix $\nu$
and $\cs$ as fluid parameters. The discretization then is $\dt \approx
\pwrr{-12} \,\mathrm{s}$ and $\dx \approx \pwrr{-9} \,\mathrm{m}.$
Again, a simulation with these parameters is not possible because a
typical pore with diameter $a\approx\pwrr{-5}\,\mathrm{m}$ would have
to be represented by $\pwrr{4}$ nodes, exceeding realistic memory
capacities. Another way to circumvent these problems is to appeal to
hydrodynamic similarity for stationary flows. The simulations in this
paper are performed with fluid parameters that represent a pseudofluid
with the same viscosity as water, but $\cs = 1 \,\mathrm{m} \,
\mathrm{s}^{-1}$ as the speed of sound. The discretization then is
$\dx= \pwrr{-6} \,\mathrm{m}$ and $\dt = \pwrr{-6} \,\mathrm{s}$.  A
pore of diameter $a$ is then represented by $10$ nodes and a cubic
sample with side-length $\pwrr{-3} \,\mathrm{m}$ requires $1000^{3}$
nodes, a manageable system size on parallel computers.  An external
force, as discussed next, drives the flow such that the velocities are
of order $\pwrr{-3}\, \mathrm{m}\,\mathrm{s}^{-1}$. The Mach and
Reynolds numbers in the simulations are $\Ma\approx\pwrr{-3}$ and
$\Re\approx\pwrr{-3}$, characterizing a laminar subsonic flow. As
long as $\Ma\ll 1$ and hydrodynamic similarity remains valid, we do
not expect that the parameters of the pseudofluid will change the
permeability estimate.

An external acceleration $\fzv(\bfl,t)$ acting on the fluid is 
implemented by adding two modifications. First, a forcing term written 
as a power series in the velocity~\cite{Ladd01}
\begin{equation}
\label{varphi:eq}
\varphi_{\ii}(\bfl,t) \!=\!  
\dt \frac{\rho \, \wwi{}}{\rhor} \! \left( h_{0} \!+\! 
\frac{ \vbf{h}_{1} \cdot\ci}{\cs^{2}} \!+\!
\frac{ \vbf{h}_{2} :(\ci\ci-\cs^{2}\vbf{I})}{2\cs^{4}}\right)\!,
\end{equation}
is added to the right hand side of Eq.~\eqref{discboltz}. Second,
Eq.~\eqref{mom_lB} for the equilibrium velocity $\vveleq$ in
Eq.~\eqref{qian} needs to be modified. 
The parameters of order 0, 1, and 2 in the expansion are $h_{0}$, 
$\vbf{h}_{1}$, and $\vbf{h}_{2}$. The definition of the velocities $\vveleq(\bfl,t)$
and $\vvel(\bfl,t)$ differ with the method used. We present four
possible implementations which all assume $h_{0}=0$, since otherwise a
source term in the mass balance would have to be taken into account.
The sums in this paragraph run from $\ii=1,\dots, N$ and the
quantities $\vbf{h}_{1}$, $\vbf{h}_{2}$, $\fii$, $\vvel$, $\vveleq$, and
$\fzv$ are functions of $\bfl$ and $t$ unless specified otherwise.

The first method to implement a body force is referred to as \mtA\ in the
remainder of the paper. It uses
\begin{equation}
\vbf{h}_{1}\! =\! \left(1\!-\!\frac{\dt}{2\tau}\right)\fzv,  \quad
\vbf{h}_{2}\! =\! \left(1\!-\!\frac{\dt}{2\tau}\right)\left(\vveleq\fzv\!+\!\fzv\vveleq\right), 
\end{equation}
and a modified definition of $\vveleq$ and $\vvel$ which causes
the influence of temporal and spatial derivatives of $\fzv$ on the density
and momentum changes to vanish. For this method one obtains $\vveleq=\vvel$, with
\begin{equation}
\label{macr:vel:A:eq}
\vvel   = \left(\sum\fii\,\ci \Big/ \sum\fii\right) + \dt\,\fzv/2
\end{equation}
instead of Eq.~\eqref{mom_lB}. A multiscale expansion in time of
the resulting discrete \LB\ equation yields that the macroscopic
density $\rho$ and velocity $\vvel$ recover the Navier-Stokes
equations with an external body force
term~\cite{PhysRevE.65.046308}. The forcing is applied in two steps
during every time step $\dt$, one half within the collision step by
the definition of $\vveleq$ and the second half within the streaming
step by the term $\varphi_{\ii}$. In the case of \LBMRT\ the part
which is applied during the collision step is added to the modes
$\m_{\ii}(\bfl,t)$ with $i=4,6,8$, which represent the momentum flux.

The second method (\mtB) is defined by setting
\begin{equation}
\vbf{h}_{1} = \fzv, \quad \vbf{h}_{2} = \vbf{0},
\end{equation}
so that $\varphi_{\ii}(\bfl)$ does not depend on
$\vveleq(\bfl,t)$. The full acceleration is applied only within the
streaming step through the term $\varphi_{\ii}(\bfl)$. One sets
\begin{equation}
\label{eql:vel:B:eq}
\vveleq = \sum\fii\,\ci \Big/ \sum\fii,
\end{equation}
and the macroscopic velocity $\vvel$ defined as in
  Eq.~\eqref{macr:vel:A:eq}.
This simplification is useful because it reduces
the computational effort, but it is restricted to
stationary flows.
In our simulations $\fzv(\bfl)$ is time independent and
we are mainly interested in the permeability and
stationary flows so that we have adopted \mtB\
in our simulations below.
In \mtB\  the macroscopic fields fulfill mass balance,
but some additional unphysical terms appear in the momentum
balance~\cite{PhysRevE.65.046308}.
Here we assume that all these additional terms
are negligible or vanish for stationary flows, 
because we expect that all spatial gradients are 
sufficiently small.

\mtC\ is intended for constant $\fzv$ and uses the same parameters
$\vbf{h}_{1}$ and $\vbf{h}_{2}$ as \mtB~\cite{bib:He97}. However, 
the macroscopic velocity $\vvel=\vveleq$ is
calculated as in Eq.~\eqref{eql:vel:B:eq}.
This recovers momentum balance, because unphysical terms either vanish or
are negligible,
but it does not recover mass balance, which in this case reads
\begin{equation}
\frac{\partial\rho}{\partial t}+\nabla\cdot(\rho\vvel)
 = -\frac{\dt}{2}\nabla\cdot(\rho\fzv).
\end{equation}
The reason is an inaccurate calculation of the macroscopic velocity
$\vvel(\bfl,t)$~\cite{PhysRevE.65.046308}. The impact of this issue
on the simulation results is shown in Sec.~\ref{diffperm:sec}.

\mtD\ suggests to incorporate the acceleration not by using the forcing term, 
but by adding the term
$\tau\fzv(\bfl,t)$ to the equilibrium velocity
$\vveleq(\bfl,t)$. The macroscopic velocity $\vvel(\bfl,t)$ remains
calculated by Eq.~\eqref{mom_lB}~\cite{SukopThorne2007}. This is equivalent to using
the forcing term with
\begin{equation}
\vbf{h}_{1} = \fzv,\quad\vbf{h}_{2} = \tau\fzv\fzv+\fzv\vvel^{*}+\vvel^{*}\fzv,
\end{equation}
and $\vvel=\vveleq$ given by Eq.~\eqref{eql:vel:B:eq}.
This implementation leads to the same drawback in the mass balance
equation as in \mtC.

The most common boundary conditions (\BC) used jointly within \LB\ implementations
are periodic (\PBC) and no-slip \BC. When using \PBC, fluid that leaves
the domain, i.e., the term $\bfl +\Delta \bfl_{\ii}$ in Eq.~\eqref{discboltz}
exceeds the computational domain size, enters the domain from the other side.
The no-slip \BC, also called simple bounce-back rule (\SBB),
approximates vanishing velocities at solid surfaces~\cite{bib:succi-01}.
If the lattice point $\bfl+\Delta\bfl_{\ii}$ in Eq.~\eqref{discboltz} 
represents a solid node, the discrete \LB\ equation is rewritten as
\begin{equation}
\label{discboltz_bbr}
\fii^{*}(\bfl,t+\dt)-\fii(\bfl,t)=\dt\,\Omega_{\ii}(\bfl,t),
\end{equation}
where the probability function $\fii^{*}$ is associated with $\ci^{*}$,
where \mbox{$\ci^{*}=-\ci$} is the probability function in 
opposite direction to $\fii$.  
Midplane \BC~\cite{SukopThorne2007} improve the
\SBB\ eliminating the zig-zag profile when plotting the mass flow $q$ vs. $\lk$, but
yield the same mass flow $Q$, see Eqs.~\eqref{massflux:q}
and~\eqref{massflux} for their definition, respectively.
The \SBB\ scheme depends on viscosity and relaxation
time $\tau$, especially in under-relaxed simulations (large values of
$\tau$)~\cite{bib:He97}. The numerically exact position of the
fluid-solid interface changes slightly for different $\tau$ which can
pose a severe problem when simulating flow within porous media, where
some channels might only be a few lattice units wide. The permeability
$\kappa$, being a material constant of the porous medium alone,
becomes dependent on the fluid viscosity. As demonstrated below within
the \LBMRT\ method this $\kappa$-$\tau$ correlation is significantly
smaller than within
\LBBGK~\cite{bib:ginzburg-dhumieres,2002RSPTA.360..437D}.
Recently, further improvements for no-slip \BC\ have been
discussed~\cite{bib:cf.CPaLLuCMi.2006}. Most of these implementations
use a spatial interpolation. For example, linearly and quadratic
interpolated bounce-back~\cite{779082,2001PF:BFL}, or
multireflection~\cite{PhysRevE.68.066614}. To calculate boundary
effects these methods use multiple nodes in the vicinity of the
surface. For this reason these schemes are unsuitable in porous media
where some pore throats might be represented by 2 or 3~nodes
only. Consequently, we use midplane \BC\ as well as \PBC\ for our
simulations.

To drive the flow on-site pressure or flux \BC~\cite{bib:pf.QZoXHe.1997,HH08b}
may be used. Using them it is possible to exactly set the ideal gas pressure (or
density, see Eq.~\eqref{p_lB}) or flux on a specific node. Thus, creating a
pressure gradient by fixing either the pressure or the mass flux at the inlet
and outlet nodes are feasible alternatives.

\section{Simulation setup} 
\label{lsetup:sec}
\begin{figure}[ht]
\begin{centering}
\includegraphics[width=0.8\columnwidth]{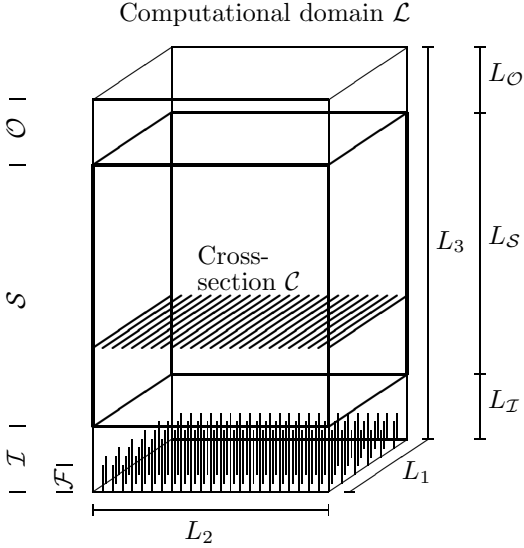}
\par\end{centering}
\caption{\label{fig:permcalc} The computational domain $\mc{L}$. The
  (porous) sample is $\mc S$, and the fluid is accelerated in the
  acceleration zone $\mc{F}$.  Two fluid chambers $\mc{I}$ and
  $\mc{O}$ are used to avoid artifacts.  }
\end{figure} 
The computational domain (see Fig.~\ref{fig:permcalc})
$\mc{L}$ is composed of three zones: the sample $\mc{S}$ 
describing the geometry
and two chambers $\mc{I}$ (inlet) and $\mc{O}$
(outlet), before and after the sample, containing fluid reservoirs.
The notation 
\begin{equation}
\label{eq:Syscross}
 \mc{C} (a):=\{\bfl \in \mc{L} : \lk=a \}
\end{equation}
denotes a cross-section, where $\Cinm$, $\Cinp$, $\Coutm$, and $\Coutp$
represent the cross-sections right before the sample
($\Cinm\in\mc{I}$), the first ($\Cinp\in\mc{S}$), and last
($\Coutm\in\mc{S}$) cross-section within the sample, and the
cross-section right after the sample ($\Coutp\in\mc{O}$),
respectively.
Every lattice point (node) in $\mc{L}$  is
either part of the matrix, denoted $\mc{M}$, or part of the
fluid, denoted $\mc{P}$, so that $\mc{M} \cup \mc{P} = \mc{L}$ 
and $\mc{M} \cap \mc{P}=\emptyset$.

Results are presented in the dimensionless quantities
\begin{equation}
\begin{split}
\hat{\x}&=\x/\dx,\quad \hat{t}=t/\dt,\quad \hat{\rho}=\rho/\rhor,\quad
\hat{p}=p/(3{\cs}^2\rhor),\\
\hat{\vvel}&=\vvel\dt/\dx,\quad \htau=\tau/\dt,\quad \htaub=\taub/\dt,\\
\hat{\kappa}&=\kappa/(\dx)^{2},\quad \hat{\fz}= \fz\, (\dt)^2/\dx,\quad
\hat{q}=q\,\dt/(\rhor\,(\dx)^{3}),
\end{split}\nonumber
\end{equation}
where the discretization parameters $\dx$ and $\dt$ are chosen
according to the analysis presented in Sec.~\ref{simmeth:sec}. Unless
otherwise noted, the relaxation time is $\htau=0.857$ and for
\LBMRT\ simulations $\htaub=1.0$ is used.
Generally, results from \LB\ simulations are labeled with the
superscript ``${\,}^{\sLB}$'', e.g., the density $\rho^{\sLB}$. If the
results refer to a specific implementation (\BGK\ or \MRT) they are
labeled accordingly, e.g., $\rho^{\sbgk}$ or $\rho^{\smrt}$.

The fluid is driven using {\sc model~B}.  The acceleration
$\fzv=\fz\,\ek $ is not applied throughout the whole domain but only
within the acceleration zone $\mc{F} \subset\mc{I}$.  An acceleration
of $\hfz=\pwrr{-6}$ is used for all simulations.

The average for a physical quantity $w$ is
\begin{equation}
\label{eq:avg}
{\langle w \rangle}_{\mc{V}}=\fracd{1}{|\mc{V}|} \sum_{\bfl \in \mc{V}}w(\bfl),
\end{equation}
with the domain $\mc{V}\in \{\mc{L}, \mc{S}, \mc{P}, \mc{I},\mc{O},\mc{F},\mc{C}(a)\}$ 
and $|\mc{V}|$ the number of nodes in that domain.
The mass flow $q$ through a cross-section $\mc{C}(a)$ is given by
\begin{equation}
\label{massflux:q}
  q(a)=\sum_{\bfl\in\mc{C}(a)\cap\mc{P}}\rho(\bfl)\,\vk(\bfl)(\dx)^{2},
\end{equation}
with $\rho(\bfl)\vk(\bfl)$ being the momentum component in direction of
the flow.
The mass flow through the whole domain is
\begin{equation}
\label{massflux}
Q=\fracd{1}{\Lk}\sum_{\lk=1}^{L_3}q(\lk).
\end{equation}

\section{Calibration} 
\label{lcalib:sec}
To calibrate the simulation we simulate Poiseuille flow in pipes with
quadratic cross-section. The simulation parameters are defined by
\begin{eqnarray}
\label{coordsystem:eq}
\tx(\bfl)&=& \left(\li-\frac{\Li+1}{2}\right)\dx\ei \nonumber\\
&&+ \left(\lj-\frac{\Lj+1}{2}\right)\dx\ej
+\left(\lk-\frac{1}{2}\right)\dx\ek, \nonumber\\
\mathcal{S}&=&\{\bfl\in\mc{L} : 2\leq \li \leq \Li-1, \, \nonumber\\
&&\qquad\ 2\leq \lj \leq \Lj-1, \,  4\leq \lk \leq \Lk-\Lout \}, \nonumber\\
\mathcal{I}&=&\{\bfl\in\mc{L}: \lk \leq \Lin\}, \nonumber\\
\mathcal{O}&=&\{\bfl\in\mc{L}: \Lk-\Lout \leq \lk \leq \Lk\}, \nonumber\\
\mathcal{F}&=&\{\bfl\in\mc{L}: \lk \leq 2 \}
\end{eqnarray}
where $\Lin=3,\Ls=\Lk-6$ and $\Lout=3$.
The system dimensions are $\Li=\Lj=\hat{B}+2,~\Lk=4\hat{B}$, with $B/\dx=\hat{B}\in
\{4,8,16,32,64\}$ the channel width.

According to Ref.~\cite{Wieg57} the analytical solution for the
velocity component in flow direction in a pipe with quadratic
cross-section is
\begin{eqnarray}
\label{vtheo}
&&{v}^{\thry}(\xi,\xj)=\lim_{M \rightarrow \infty}{v}(\xi,\xj,M) \\
&&{v}(\xi,\xj,M)=-\frac{\dpl}{2\eta}\left( \frac{B^{2}}{4}-{\xj}^{2}
-\frac{8B^{2}}{\pi^{3}} \sum_{n=0}^{M}C_{n}\right),\nonumber\\
&&C_{n}=(-1)^{n}\frac{\cosh\left(\fracd{(2n+1)\pi}{B}\xi\right)
\cos\left(\fracd{(2n+1)\pi}{B}\xj\right)}
{(2n+1)^{3}\cosh\left(\fracd{(2n+1)\pi}{2}\right)}\nonumber
\end{eqnarray}
where $\xi\in[-B/2,B/2],\xj\in[-B/2,B/2]$.
The Cartesian coordinates $\xi$ and $\xj$ have
their origin in the center of the pipe. 
$\dpl$ is the pressure gradient in
flow direction and $\eta$ the dynamic viscosity.
The expression $v(\xi,\xj,M)$ is asymmetric in $\xi$ and $\xj$. Contrary to the
no-slip condition the velocities $v(B/2,\xj,M)$ are not zero for
finite $M$. To estimate the truncation error we define
\begin{equation}
\label{vnorm}
 \tilde{v}(x,M) = \frac{2\eta}{\dpl B^2}\,{v}(B/2,x,M),\quad x\in[-B/2,B/2],\\
\end{equation}
and
\begin{equation}
\label{vwall}
\left\Vert {\tilde{v}_{\rm wall}}(M)\right\Vert_{2}:=\sqrt{\frac{1}{B} \int_{-B/2}^{\phantom{-}B/2}\ \left| 
\tilde{v}(x,M) \right|^{2}\,dx },
\end{equation}
with $\tilde{v}(x,M)$ being the normalized velocities on the wall
calculated from Eq.~\eqref{vnorm}. $\left\Vert
{\tilde{v}_{\rm wall}}(M)\right\Vert_{2}$ quantifies the truncation error at finite
$M$. Requiring that the truncation error
$\left\Vert \tilde{v}_{\mathrm{wall}} (M)\right\Vert_{2}$ is at least
three to four decades smaller than the velocities in the corners, for
example $\tilde{v}(\tx(1,1,L_3/2)_1,\tx(1,1,L_3/2)_2,M)$
or any other such corner velocity, yields $M\approx 50$.
For all further comparisons with \LB\
simulations we use $M=200$.
If $M$ is chosen too small, a meaningful comparison of the simulation results
with the analytical solution is not possible because of the inaccuracies in
the numerical evaluation of the analytical solution itself.

Eq.~\eqref{vtheo} is the stationary solution for the velocity
component in flow direction on a quadratic cross-section in an infinitely
long pipe and for a constant pressure gradient. Therefore, the
simulated $v^{\sbgk}(\bfl,t)$ and $\rho^{\sbgk}(\bfl,t)$ are inspected
for convergence at the end of the simulation $t = t_{\mathrm{end}}$
and the assumption of a constant pressure gradient is checked. We
define
\begin{equation}
\label{deltav}
\delta w(t,dt)=\underset{\bfl\in (\mc{S}\cap\mc{P})}
\max \left(\frac{{w}(\bfl,t)-{w}(\bfl,t-dt)}
{{w}(\bfl,t)}\right),
\end{equation}
as the maximum relative change of a quantity $w$ during the time $dt$
and within the computational domain $\mc{S}\cap\mc{P}$, where
${w}(\bfl,t)$ is either the velocity $v^{\sbgk}(\bfl,t)$ or the
density ${\rho}^{\sbgk}(\bfl,t)$. Because the
pressure is proportional to the density, Eq.~\eqref{p_lB}, the
pressure is converged, if the density is sufficiently converged.
\begin{table}[b]
\begin{centering}
\begin{tabular}{|c|c|c|c|c|c|}
\hline $\hat{B}$ & $dt/\dt$ &$\tend/\dt$ & $\delta
v(\tend,dt)$ & $\delta\!\rho(\tend,dt)$
\tabularnewline &$[\pwr{3}]$ &$[\pwr{3}]$ & $[\pwr{-8}]$ &
$[\pwr{-4}]$ \tabularnewline \hline \hline 4 & 1 & 20 & 2.34 & 0.174
\tabularnewline \hline 8 & 1 & 20 & 2.80 & 0.174 \tabularnewline
\hline 16 & 5 & 30 & 4.27 & 0.869 \tabularnewline \hline 32 & 5 & 50 &
7.45 & 0.869 \tabularnewline \hline 64 & 10 & 120 & 1.33 & 1.74
\tabularnewline \hline
\end{tabular}
\end{centering}
\caption{\label{tbl:steady} Maximum relative change of the velocity
  $\delta v(\tend,dt)$ and density
  $\delta\!\rho(\tend,dt)$, Eq.~\eqref{deltav},
  during the time $dt$ when the simulation ended at
  $\tend$.  $\hat{B}$ is the dimensionless channel width.}
\end{table}
The results from Eq.~\eqref{deltav} are shown in Tab.~\ref{tbl:steady}.  In the
simulations the velocities are of order $\hat{v}^{\sbgk}(\bfl,t) \approx
\pwrr{-4}$ so the absolute changes are of order $\pwrr{-12}$, using the relative
changes $\delta {v}$ from Tab.~\ref{tbl:steady}.  The fluid density is
$\hat{\rho}^{\sbgk}(\bfl,t) \approx 1.0$ giving absolute changes of order
$\pwrr{-5}$. The variation of the pressure gradient can be approximated by
$2\,\delta\!\rho/(L_3) < \pwrr{-7}$.  When calculating errors by comparing them
with analytical solutions the number of significant digits is determined by the
convergence of the simulation.  We use the notation ${v}^{\sbgk}(\bfl)$,
${\rho}^{\sbgk}(\bfl)$ and ${p}^{\sbgk}(\bfl)$ for the velocity, density and
pressure at the end of the simulation $t=t_{\mathrm{end}}$.

Due to the way we drive the flow, the pressure increases in the
  acceleration zone $\mc{F}$ and then decreases along the flow
  direction. See Fig.~\ref{rho_fig}, where the average density
${\langle\hat{\rho}\rangle}_{\mc{C}(\lk)\cap\mc{P}}-1$, from a
\LBBGK\ simulation for a pipe of width $\hat{B}=7$ is shown. To verify that
the pressure gradient $\dpl$ can be assumed to be constant as required
by Eq.~\eqref{vtheo}, we linearly fit $\langle
p^{\sbgk}\rangle_{\mc{C}(\lk)\cap\mc{P}}$ inside the sample.  In the
\LB\ simulations, for pipes of width $\hat{B}=4$, $8$, $16$, $32$, and $64$,
all residues of the linear fit are of the order $1\pwr{-9}$, so that
the pressure gradient can be assumed to be constant.

\begin{figure}[h]
\begin{centering}
\includegraphics[width=1.0\columnwidth]{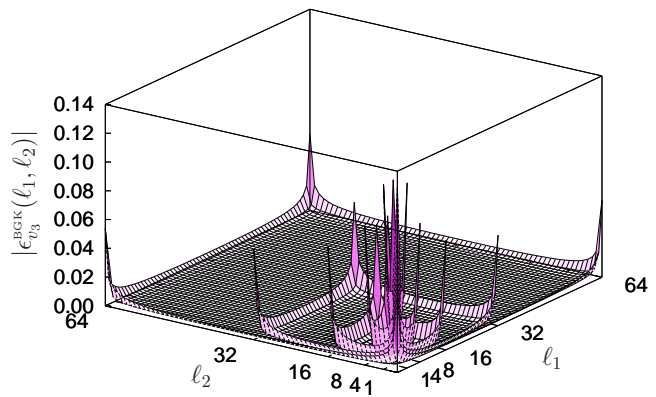}
\end{centering}
\caption{\label{fig:relerr3d} Overview of the relative error
  $|\epsilon^{\sbgk}_{v}(\bfl)|$ with $\lk=\Lk/2$, for
  pipes of widths $\hat{B}\in \{4,8,16,32,64\}$.  Nodes at the corners cause
  the largest error followed by those close to the solid walls. For larger pipes
  the error decreases substantially.}
\end{figure}

\begin{figure}[h]
\begin{centering}
\includegraphics[width=0.9\columnwidth]{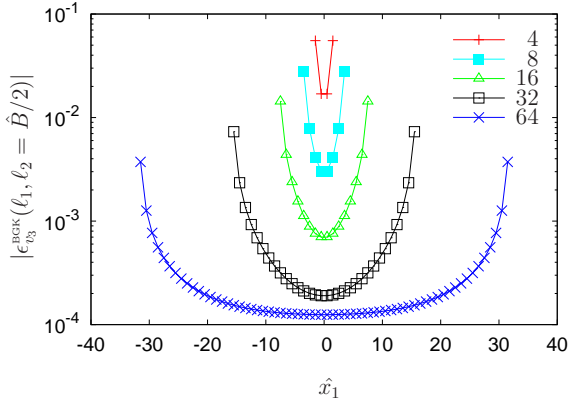}
\end{centering}
\caption{\label{fig:relerr_cline} Log-linear plot of the relative
  error along the central line on a cross-section, i.e
  $|\epsilon^{\sbgk}_{v}(\bfl)|$ with $\lj=\Lj/2$ and $\lk=\Lk/2$. The
  error is largest at the walls and declines towards the center of the
  pipe. Different line styles indicate different pipe widths
$\hat{B}=4,8,16,32,64$ as shown in the legend.}
\end{figure}

\begin{figure}[h]
\begin{centering}
\centerline{\BGK}
\includegraphics[width=0.9\columnwidth]{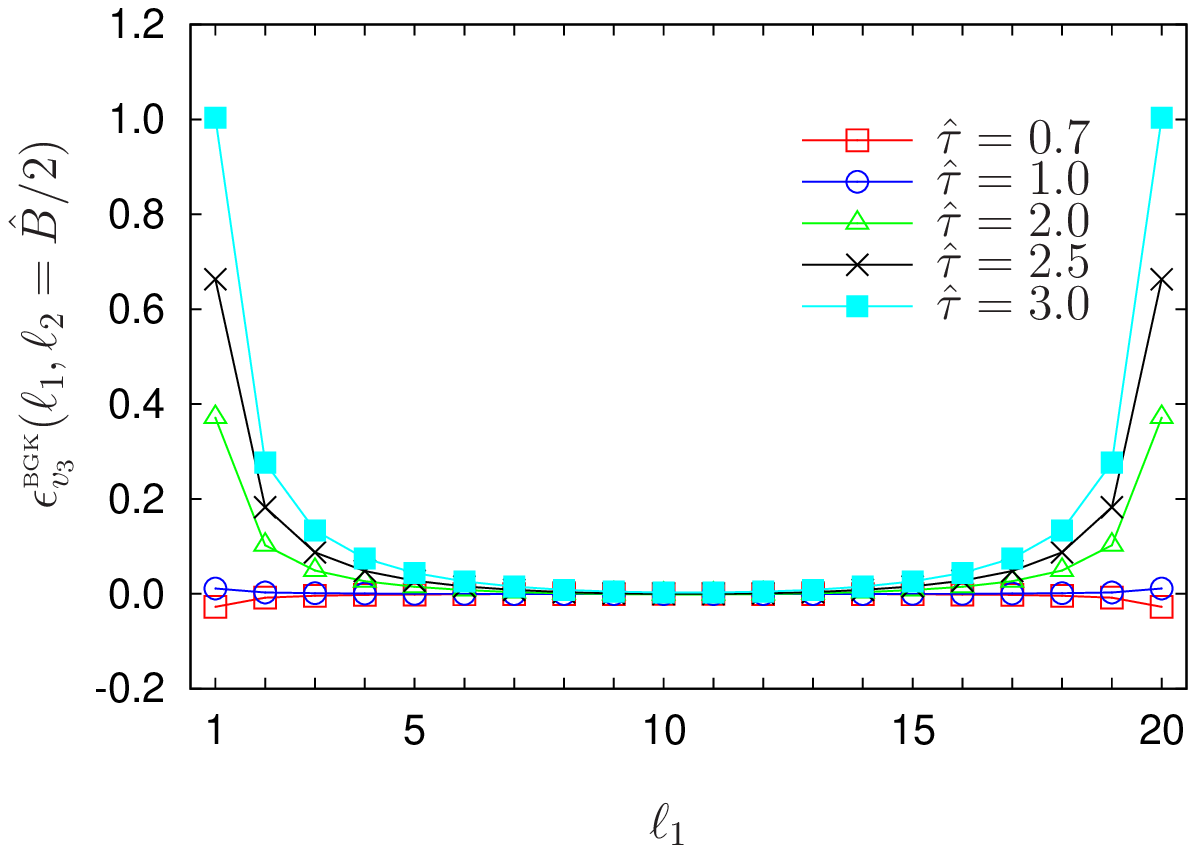}
\centerline{\MRT}
\includegraphics[width=0.9\columnwidth]{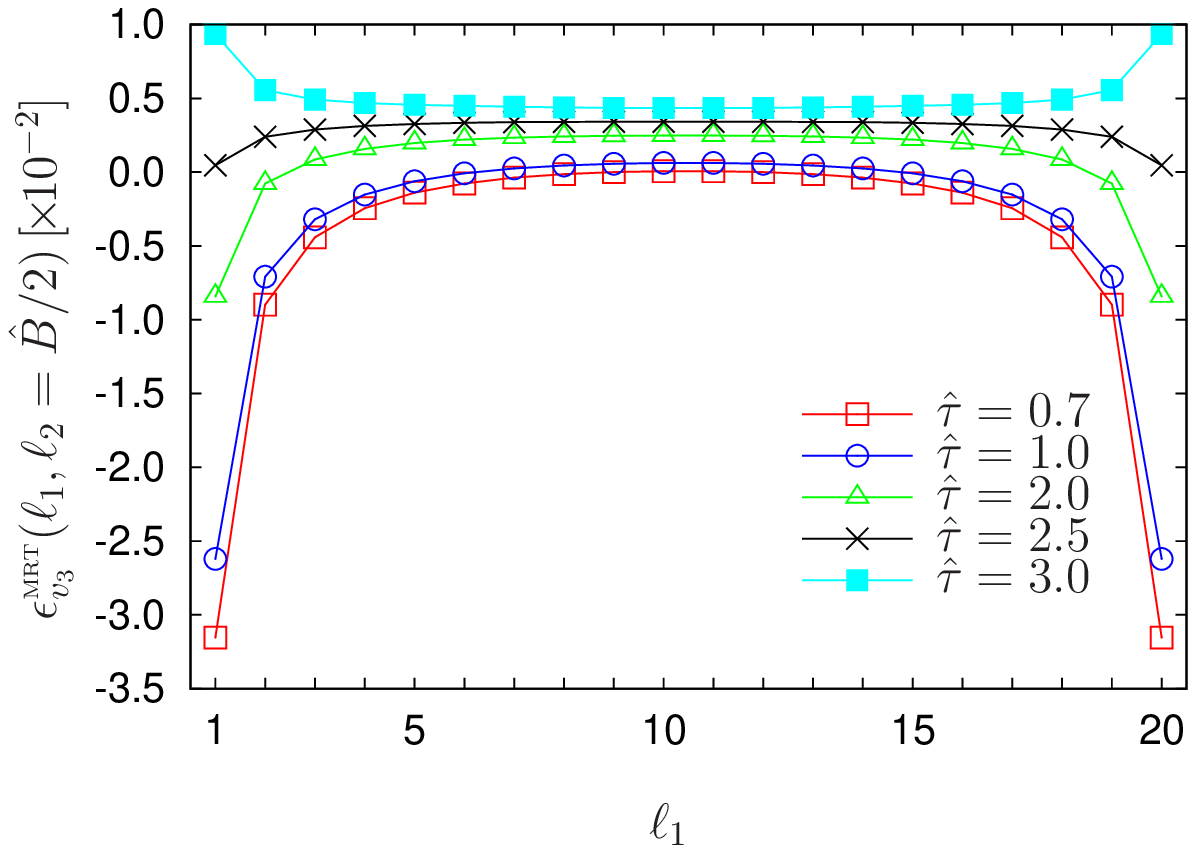}
\end{centering}
\caption{\label{plot_3d_a20} Relative error $\epsilon^{\sLB}_{v}(\bfl)$ with
  $\lj=\hat{B}/2$ and $\lk=\Lk/2$ for flow in a quadratic pipe of width
$\hat{B}=20$ and different
  values of $\htau$. The upper figure shows the results for the \BGK\
  implementation and in the lower figure the \MRT\ results are shown. The
  relative error of the \MRT\ simulations is smaller by $\pwrr{-2}$ than the
  \BGK\ results.}
\end{figure}

Next, the velocity component in flow direction $v^{\sLB}(\bfl)$ with
$\bfl \in \mc{C}(\Lk/2)$ is compared to the analytical solution
Eq.~\eqref{vtheo}, evaluated at the node positions
${v}^{\thry}(\tx(\bfl)_{\ai},\tx(\bfl)_{\aj})$.  The cross-section
$\mc{C}(\Lk/2)$ is chosen to minimize finite size effects and
artifacts from the in/outlet chamber.  We define absolute and relative
errors of the velocities as
\begin{equation}
\label{vabserr}
e^{\sLB}_{v}(\bfl):=v^{\sLB}(\bfl)-v^{\thry}(\tx(\bfl)_{\ai},\tx(\bfl)_{\aj}),
\end{equation}
\begin{equation}
\label{vrelerr}
\epsilon^{\sLB}_{v}(\bfl):=\frac{v^{\sLB}(\bfl)-
v^{\thry}(\tx(\bfl)_{\ai},\tx(\bfl)_{\aj})}{v^{\thry}(\tx(\bfl)_{\ai},\tx(\bfl)_{\aj})}.
\end{equation} 
Fig.~\ref{fig:relerr3d} provides an overview on the structure of
$|\epsilon^{\sbgk}_{v}(\bfl)|$ and Fig.~\ref{fig:relerr_cline} shows
$|\epsilon^{\sbgk}_{v}(\bfl)|$ with $\lj=\Lj/2$ and $\lk=\Lk/2$ as a
log-linear plot for different pipes of width $\hat{B}$ and \LBBGK.  The
largest relative errors are located in the corners and close to the
wall. As the resolution increases the relative error declines
rapidly.  In the central region it is much smaller than 1\%.
To gain insight into how strong the value of the relaxation-time
$\tau$ influences the accuracy of the velocity field, simulations with
\LBBGK\ and \LBMRT\ and different relaxation-times $\htau=0.7$, $1.0$,
$2.0$, $2.5$, and $3.0$ are investigated.  Fig.~\ref{plot_3d_a20}
displays the relative error of the velocity
$\epsilon^{\sLB}_{v}(\bfl)$, Eq.~\eqref{vrelerr}, with $\lj=\hat{B}/2$,
$\lk=\Lk/2$, and $\hat{B}=20$ for both implementations \LBBGK\ and
\LBMRT\ and different relaxation times. The calculated velocity in the
center of the pipe is in good agreement with the theoretical solution,
having a relative error smaller than $1\%$.  It is interesting to note
that when using the \LBBGK\ method the largest error occurs for a
large relaxation time $\htau=3.0$ (over relaxation), whereas the
largest error for the \LBMRT\ result occurs at a small relaxation time
$\htau=0.7$ (under relaxation). The calculated velocities tend to be
overestimated $\epsilon^{\sbgk}_{v}(\bfl)>0$ for \LBBGK\ simulations
and underestimated $\epsilon^{\smrt}_{v}(\bfl)<0$ for
\LBMRT\ simulations. When using \LBMRT\ the relative error is smaller
by roughly a factor $\pwrr{-2}$ when compared to results of the
\LBBGK\ method.

For permeability calculations from Darcy's law, see Eq.~\eqref{darcy}, the mean
velocity $\langle v^{\sLB}\rangle_{\mc{S}}$ is used.  Therefore, the mean
relative error $\langle|\epsilon^{\sbgk}_{v}|\rangle_{\mc{S}}$ and the mean
absolute error $\langle |e^{\sbgk}_{v}|\rangle_{\mc{S}}$ are of interest. Both
decrease when $\hat{B}$ increases, as shown in Tab.~\ref{tbl:rel_err_v}. The mean
relative error shows a power law behavior $ \langle | \epsilon^{\sbgk}_{v}|
\rangle_{\mc{S}}=a_1\hat{B}^{a_2} $, with parameters $a_1\approx0.6$ and
$a_2\approx-1.6$. This relation can be used to calculate the relative error for
arbitrary pipe widths $B$.  Overall, the \LBBGK\ implementation is able to
reproduce the velocity field for quadratic pipes very accurately. The mean
relative error $ \langle | \epsilon^{\sbgk}_{v}| \rangle_{\mc{S}}$ is below 1\%
if the pipes are resolved better than $\hat{B}>14$.
\begin{table}[b]
\begin{centering}
\begin{tabular}{|c|c|c|} 
\hline 
$\hat{B}$ & ${\langle |\epsilon^{\sbgk}_{v}|\rangle}_{\mc{S}} $ 
& ${\langle |\hat{e}^{\sbgk}_{v}| \rangle}_{\mc{S}}\, [\pwr{-6}]$ \tabularnewline 
  \hline 
  \hline 
4 & 0.064099 & 0.300    \tabularnewline 
\hline 
8 & 0.021840 & 0.114    \tabularnewline 
\hline 
16 & 0.007241 & 0.068   \tabularnewline 
\hline 
32 & 0.002331 & 0.034   \tabularnewline 
\hline 
64 & 0.000784 &0.023    \tabularnewline \hline
\end{tabular}
\end{centering}
\caption{\label{tbl:rel_err_v} Mean relative error $\langle
|\epsilon^{\sbgk}_{v}| \rangle_{\mc{S}}$ and  absolute error $\langle|
\hat{e}^{\sbgk}_{v}| \rangle_{\mc{S}}$ for different pipes with width
$\hat{B}$. The
error declines rapidly as the pipe width increases.}
\end{table}

\begin{figure}[ht]
\begin{centering}
\includegraphics[width=0.9\columnwidth]{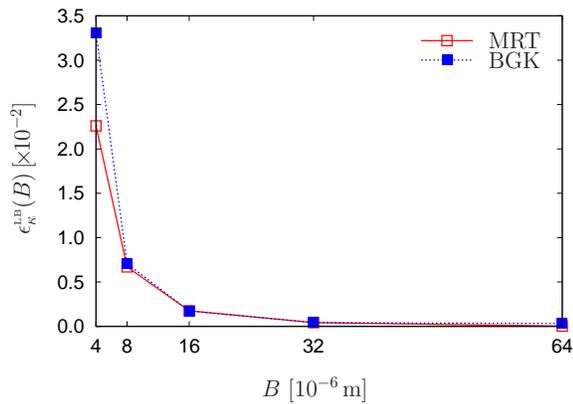}
\end{centering}
\caption{\label{fig:mrtvssrt2} Relative error $\epsilon_{\kappa}(B)$,
  of the permeability $\kappa^{\sLB}(B)$ versus channel width $B$ at fixed
resolution
$\dx= 10^{-6}$m as
  calculated using \LBBGK\ (solid line) and \LBMRT\ (dashed line)
  simulations.}
\end{figure}

Following the evaluation of the calculated velocity field,
permeabilities are calculated using both implementations \LBBGK\ and
\LBMRT. The permeability $\kappa^{\sLB}$ is calculated using Darcy's
law:
\begin{equation}
\label{darcy}
\kappa^{\sLB} = -\eta \frac{{\langle v^{\sLB} \rangle}_{\mc{S}}}
{{\langle \dpl^{\sLB}  \rangle}_{\mc{S}}},
\end{equation}
where $\eta$ is the dynamic viscosity, ${\langle \vk^{\sLB}
  \rangle}_{\mc{S}}$ is the average velocity in the sample and
${\langle\dpl^{\sLB} \rangle}_{\mc{S}}$ is the average pressure
gradient component in direction of the flow. Details how
${\langle\dpl^{\sLB}\rangle}_{\mc{S}}$ can be determined from
$\rho^{\sLB}(\bfl)$ will be discussed later in this article. The
dynamic viscosity is calculated as $\eta=\nu\bar{\rho}^{\sLB}$ with
$\bar{\rho}^{\sLB}={\langle\rho^{\sLB}\rangle}_{\mc{P}\cap \mc{S}}$
approximated by
\begin{equation}
\label{rhop}
\bar{\rho}^{\sLB} \approx
\frac{{\langle\rho^{\sLB}\rangle}_{\Cinp}+{\langle\rho^{\sLB}\rangle}_{\Coutm}}{2}.
\end {equation}
The analytically obtained permeability is~\cite{{Wieg57}}
\begin{equation}
\label{k_square_th}
\!\kappa^{\thry}(B)\!=\!\lim_{M \rightarrow \infty}\!\frac{B^{2}}{4}\!\left(\!
  \frac{1}{3}\!-\!\frac{64}{\pi^{5}}\!
  \sum_{n=0}^{M}\!\frac{\tanh\left((2n\!+\!1)\fracd{\pi}{2}\right)}{(2n\!+\!1)^{5}}
\!\right)\!,
\end{equation} 
where we use $M=200$ for numerical evaluation.  To evaluate the error
we define
\begin{equation}
\label{krelerr}
\epsilon^{\sLB}_{\kappa}(B):=\frac{\kappa^{\sLB}(B)-\kappa^{\thry}(B)}{\kappa^{\thry}(B)}.
\end{equation} 
The relative errors $\epsilon^{\sbgk}_{\kappa}(B)$ and
$\epsilon^{\smrt}_{\kappa}(B)$ are shown in Fig.~\ref{fig:mrtvssrt2}
and it can be observed that they fall below 1\% for all pipes wider
than $B=16\dx$.  It seems that the \LBBGK\ method is slightly more
accurate, but the relaxation time $\htau=0.857$ was 
fine tuned to reproduce the exact result with \LBBGK.
Fig.~\ref{fig:mrtvssrt2} shows that an
adjusted relaxation parameter $\htau$ can make up for the methodically
inferior \LBBGK\ implementation.  In realistic porous media, however
it is not possible to determine an optimal relaxation time $\tau$,
because the pore diameters and pore throats vary, although a useful
range of $\tau$ can be determined, see
Sec.~\ref{diffperm:sec}. Therefore the \LBMRT\ method is more reliable
as its results are less dependent on $\tau$.

The results for the velocity field and permeability show that even for
a simple quadratic channel a resolution of at least $20$ lattice nodes is
required to achieve an accuracy of the permeability of order $1\%$. At
present, discretization at this resolution is neither experimentally
available nor computationally manageable.

\section{Potential difficulties leading to inaccuracies}
\label{diffperm:sec}
In this section we discuss typical difficulties arising when
calculating permeabilities for complex geometries. This includes the
influence of the relaxation time $\tau$ on the permeability, the
accurate approximation of the average pressure gradient, the
implementation of the external force and the discretization error.

When using \SBB, the relaxation time $\tau$ slightly changes the
position of the boundary between adjoined fluid-solid nodes. Due to
this effect the relaxation time has a substantial influence on the
permeability calculation~\cite{bib:cf.CPaLLuCMi.2006}.  One also has
to be aware that this effect is always correlated with the
discretization error and cannot be corrected analytically when
investigating stochastic porous media.
\begin{figure}[ht]
   \begin{centering}
 \includegraphics[width=0.9\columnwidth]{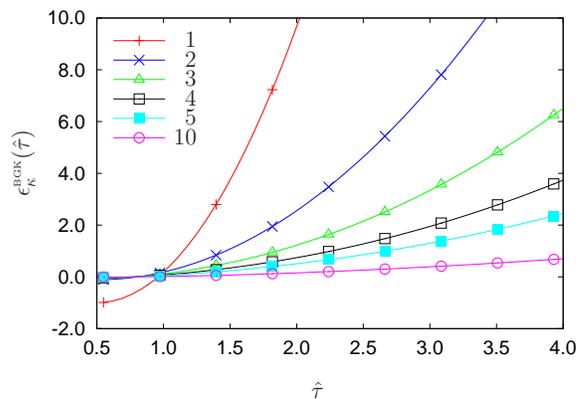}
   \end{centering}
   \caption{\label{permtau_fig}Relative error $\epsilon^{\sLB}_{\kappa}$ 
     vs. the value of $\htau$ for a Poiseuille flow in a quadratic pipe with 
     different pipe width $\hat{B}=1,2,3,4,5,10$ as indicated by different
line styles.}
\end{figure}
\begin{figure}[ht]
\hfill
 \includegraphics[width=0.42\columnwidth]{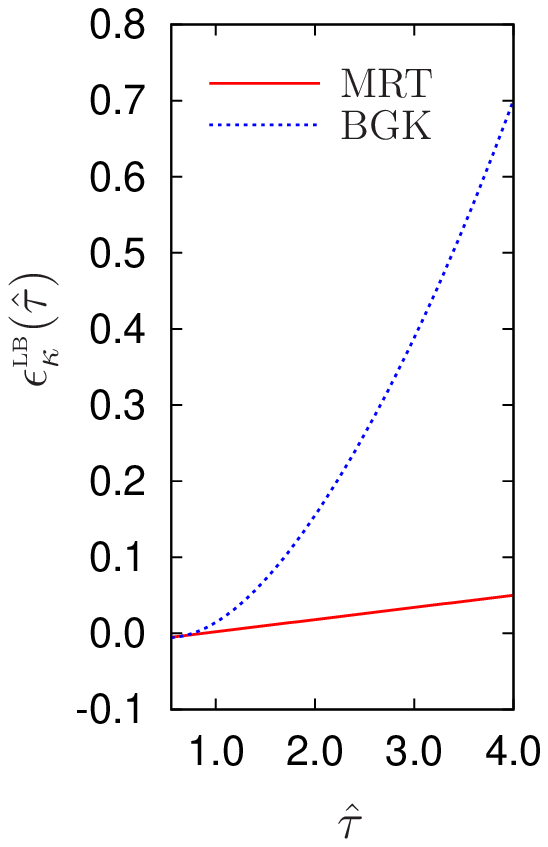}
\hfill
 \includegraphics[width=0.42\columnwidth]{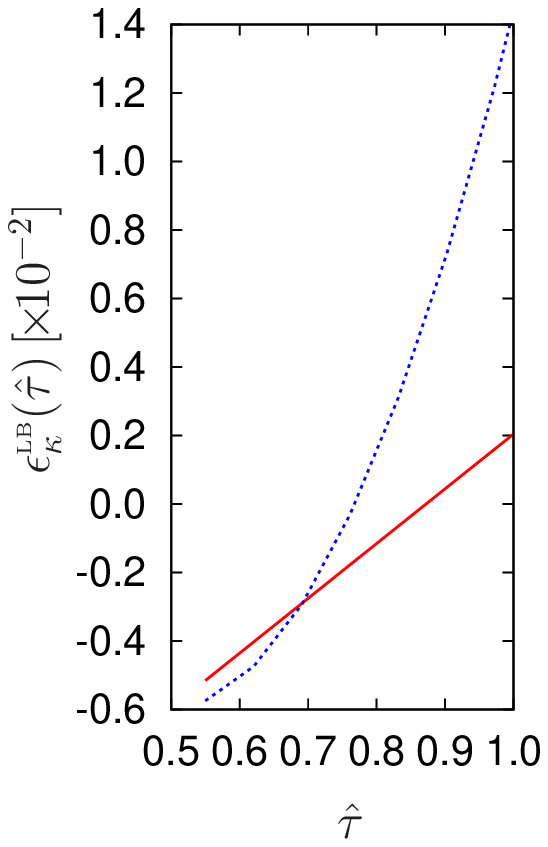}
\hfill
   \caption{\label{permtau_bgkmrt_fig}Relative error 
     $\epsilon^{\sLB}_{\kappa}$ for 
     Poiseuille flow in a quadratic pipe with width $\hat{B}=10$, 
     calculated using \LBBGK, dashed line, and \LBMRT,
     solid line. On the left, the interval
     $\htau\in[0.5,4.0]$ is shown and on the right, a zoom 
     in the interval $\htau\in[0.5,1.0]$ can be seen.}
\end{figure}
To analyze the influence of $\tau$ on the permeability we again
investigate Poiseuille flow in quadratic pipes using a computational grid
aligned with the pipe geometry to minimize the discretization
error. The relative error $\epsilon^{\sLB}_{\kappa}(\htau)$ is shown in Figs.~\ref{permtau_fig}
and~\ref{permtau_bgkmrt_fig}. For over-relaxed systems ($\htau>1$),
the \LBBGK\ method yields incorrect results, increasing dramatically
the dependence of permeability on $\tau$ when the geometry is poorly
discretized ($\hat{B} \le 5$). If $\htau\in[0.5,1.0]$, the absolute error of
permeability estimation is less than 3\% for all $\hat{B}>2$. The
\LBMRT\ method has to be considered more reliable in general because
the influence of $\tau$ is much smaller and the influence of $\taub$
is practically insignificant. For example, in
Fig~\ref{permtau_bgkmrt_fig}, using a value $\htau=3.5$, the error of
the \LBBGK\ method is 53.70\%, while the error of the \LBMRT\ method
is 4.213\%.  However, if only a small number of nodes is used ($\hat{B} \ge
5$) even the \LBMRT\ method produces a substantial error.
Fig.~\ref{permtau_bgkmrt_fig} compares the results between the
\LBBGK\ and \LBMRT\ for a pipe with $\hat{B}=10$. Here, both absolute errors
in the interval $\htau\in[0.5,1.0]$ are smaller than 1.5\%.  It is
important to stress here that outside the interval $\htau\in[0.5,1.0]$
the \LBMRT\ results remain accurate when $B$ decreases, which is not
the case for \LBBGK.

\begin{figure}[ht]
   \begin{centering}
\includegraphics[width=0.9\columnwidth]{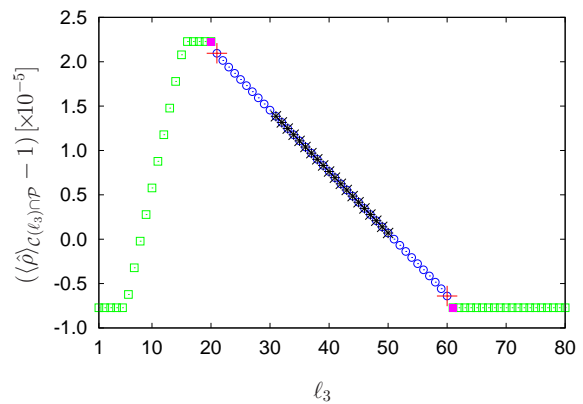}
   \end{centering}
   \caption{
     Average density $({\langle\hat{\rho}\rangle}_{\mc{C}(\lk)\cap\mc{P}}-1)$ 
     vs. $\lk$ as obtained from a \LBBGK\ simulation. The sample $\mc{S}$ 
     is placed in $\lk\in[21,60]$, leaving 20 nodes
     before and after the sample as injection chamber. 
     Within the acceleration zone $\mc{F}$ at $\lk\in[6,15]$ the density
and pressure increase.
     Alternatives to measure the
     density or pressure gradient: a)~Use all nodes inside the sample for a linear fit ($\odot$). 
     b)~Linear fit not using the nodes closer than $W=10$ to the point 
     where the fluid enters 
     or leaves the sample ($\times$). 
     c)~Use the first and last cross-section $\Cinp$ and $\Coutm$ inside the sample
     ($+$). 
     d)~Use the cross-section $\Cinm$ and $\Coutp$ 
     ($\blacksquare$). The nodes represented by $\square$ define 
     the injection channel ($\mc{I}$ and $\mc{O}$).}
\label{rho_fig}
\end{figure}

To compute the permeability using Darcy's law as given in Eq.~\eqref{darcy},
the average pressure gradient in direction of the flow ${\langle \dpl
\rangle}_{\mc{S}}$ is required. Because the permeability can strongly
depend on the way the pressure gradient is obtained, alternative methods
for its determination are discussed:
\begin{enumerate}[a)]
\item Calculating the slope of a linear fit through 
the full data set ${\langle p \rangle}_{\mc{C}(\lk)\cap \mc{P}}$ 
obtained using all cross-sections $\mc{C}(\lk)$, $\Lin+1\leq\lk\leq\Lk-\Lout$.
\item As a), but using only the cross-sections $\mc{C}(\lk)$,
$\Lin+1+W\leq\lk\leq\Lk-\Lout-W$, $W\in\mathbbm{N} $, see Fig.~\ref{rho_fig}. 
The cross-sections closer than $W$ to the inlet and outlet of the sample are
not taken into account.  The idea is to minimize boundary effects.
\item Approximation of ${\langle\dpl\rangle}_{\mc{S}}$ by the arithmetic mean of the
  pressure at $\Cinp$ and $\Coutm$, i.e.,
\begin{equation}
\label{dpp}
 {\langle \dpl \rangle}_{\mc{S}} 
 \approx \frac{{\langle p \rangle}_{\Coutm\cap \mc{P}}-{\langle p \rangle}_{\Cinp\cap \mc{P}}}
 {(\Ls-1)\dx},
\end{equation}
where $\Ls$ is the sample length.
\item Approximation of ${\langle\dpl\rangle}_{\mc{S}}$ by the arithmetic mean of the
  pressure at $\Cinm$ and $\Coutp$, i.e.,
\begin{equation}
 {\langle \dpl \rangle}_{\mc{S}} 
 \approx \frac{{\langle p \rangle}_{\Coutp\cap \mc{P}}-{\langle p \rangle}_{\Cinm\cap \mc{P}}}
 {(\Ls+1)\dx}.
\end{equation}
\end{enumerate}
For a quantitative comparison of the different methods simulations
with the following parameters are performed:
\begin{equation}
\label{domain:eq}
\begin{split}
\Li&=40,\,\Lj=40,\,\Lk=80, \\
\mc{S}&=\{\bfl\in\mc{L} : 17 \leq \li \leq 23 , 
17 \leq \li \leq 23 ,
21\leq \lk \leq 60\}, \\
\mc{I}&=\{\bfl\in\mc{L} : 1\leq \lk \leq 20\}, 
\mc{O}=\{\bfl\in\mc{L} : 61\leq \lk \leq 80\}, \\
\mc{F}&=\{\bfl\in\mc{L} : 6 \leq \lk \leq 15\}, 
\Lin=20,\Ls=40,\Lout=20.\nonumber
\end{split}
\end{equation}
The average density
${\langle\hat{\rho}\rangle}_{\mc{C}(\lk)\cap\mc{P}}$, is shown in
Fig.~\ref{rho_fig}.  Although the density field $\rho(\bfl)$ is
continuous, the average ${\langle \rho \rangle}_{\mc{C}(\lk)\cap\mc{P}}$
shows two discontinuities, one at the beginning of the sample
($\lk=20$) and one at the end of the sample ($\lk=60$). These can be
explained by the small compressibility of the fluid. The majority of
the fluid in chamber $\mc{I}$ flows towards the surface of the sample
causing an increased local density. The same effect can be observed
right behind the sample where one finds a low density due to the
fluid compressibility. Because we use periodic boundary conditions, the
pressure is almost constant in both chambers $\mc{I}$ and $\mc{O}$ and
only increases in the acceleration zone $\mc{F}$. The small increment
right before the sample and the small decrement right after the sample
are both imperceptible in Fig.~\ref{rho_fig}.
The results presented in Tab.~\ref{tbl:pdl} show that by using
alternatives a), b) and c) an error smaller than $1\%$ can be
obtained. The remaining method d), however, shows a substantially
larger error and is therefore not suitable for measuring the pressure
gradient.  Alternative b) is not taking into account the
cross-sections closer than $W=10$ to the inlet and outlet of the
sample. Changing $W$ does not influence the accuracy much.  For stochastic
porous media we suggest to use alternative~c) because it is very easy 
to implement and no fit is necessary.
The last row of Tab.~\ref{tbl:pdl} shows the values
obtained without an injection chamber and with a force acting
throughout the whole domain. Even though the result is
accurate, this method has a major disadvantage, because it can only
be applied to periodic samples and not to stochastic porous media. In
realistic porous media, chambers before and after the sample are
necessary to provide a fluid reservoir but they might
decrease the accuracy of the method due to disturbances of the
velocity field at the in- and outlet.
\begin{table}[htb]
\begin{tabular}{|l|c|c|c|c|}
\hline 
$\hat{B}=7$ & a) & b) & c) & d) \tabularnewline
\hline
\hline 
$\hat{\kappa}^{\sLB}$ & 1.73244 & 1.73689 & 1.71346 &  1.64270 \tabularnewline
\hline 
$\epsilon_{\kappa}\,[\%]$ & 0.61423 & 0.87278 & 0.48787 & 4.59761 \tabularnewline
\hline \hline
\multicolumn{5}{|c|}{Without injection channel} \tabularnewline
\hline \hline 
$\hat{\kappa}^{\sLB}$ & \multicolumn{4}{|c|}{1.73689}  \tabularnewline
\hline 
$\epsilon_{\kappa}\,[\%]$ & \multicolumn{4}{|c|}{0.87277} \tabularnewline
\hline
\end{tabular}
\caption{Results of the three alternatives to measure
${\langle\dpl\rangle}_{\mc{S}}$ and the case without using an injection
channel. Shown are calculated permeabilities and their relative error for
$\hat{B}=7$. For alternative b) $W=10$ is used.}
\label{tbl:pdl}
\end{table}

An important point for performing high precision permeability
measurements is the way the pressure gradient is generated. While
pressure boundary conditions provide a well defined way of fixing the
pressure at the in- and outlet, they assume an ideal gas and 
are slightly harder to implement than a simple body force driving the flow. 
In addition, even though
the pressure is fixed before and after the sample, an injection
chamber is still required and for high precision permeability
measurements one has to measure the pressure gradient as discussed
above. Therefore, most \LB\ implementations found in the literature
use body forces. In fact, all papers
we are aware of, that have been published before 2002, and a large
fraction of more recent publications use an incorrect force
implementation which can lead to severely erroneous
permeabilities. Popular examples for such implementations are \mtC\
and \mtD. They lead to an underestimation of the velocity
$\vvel(\bfl)$ in the direction of the flow on the lattice nodes where
the acceleration is acting.  Many publications apply the force  throughout 
the whole simulation domain. The results obtained from such implementations cannot be
trusted for two reasons: Firstly,
in \mtC\ and \mtD\  
the macroscopic velocity in the acceleration zone 
is smaller than the correct value. In some cases it can even 
be negative.
Secondly, the pore structure plays an important role. The number of nodes at any
cross-section $\mc{C}(\lk)$ determines the number of times the
additional acceleration term has to be added to the mass flow in order
to assure a constant flux.  

\begin{figure}[ht]
   \begin{centering}
 \includegraphics[width=0.9\columnwidth]{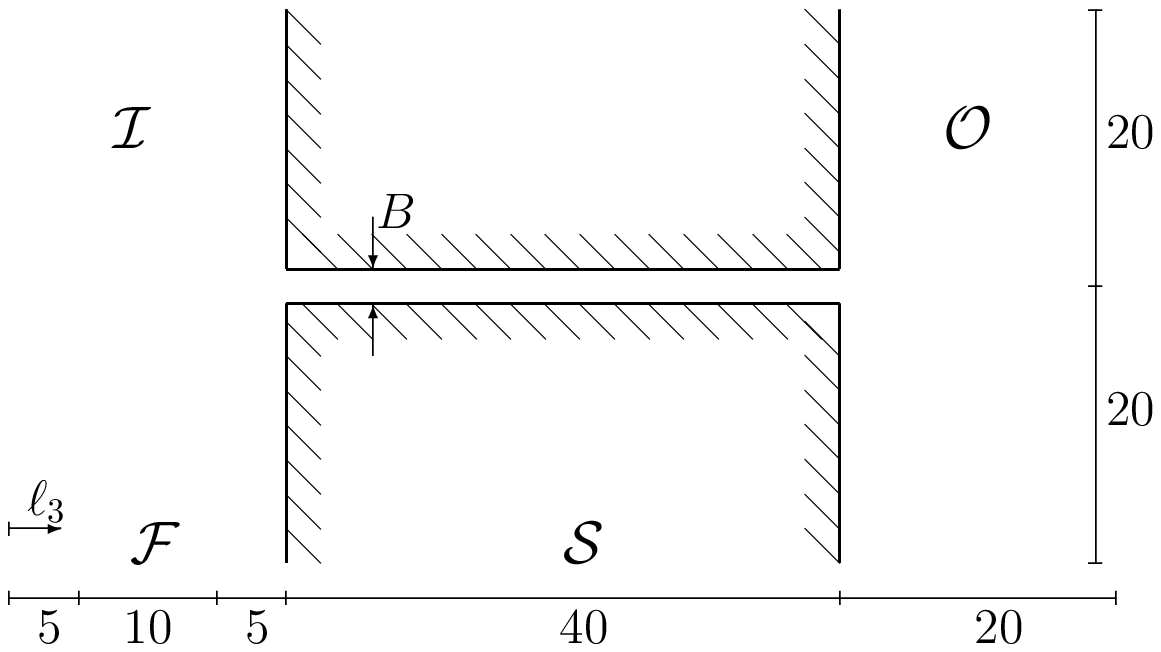}
 \includegraphics[width=0.9\columnwidth]{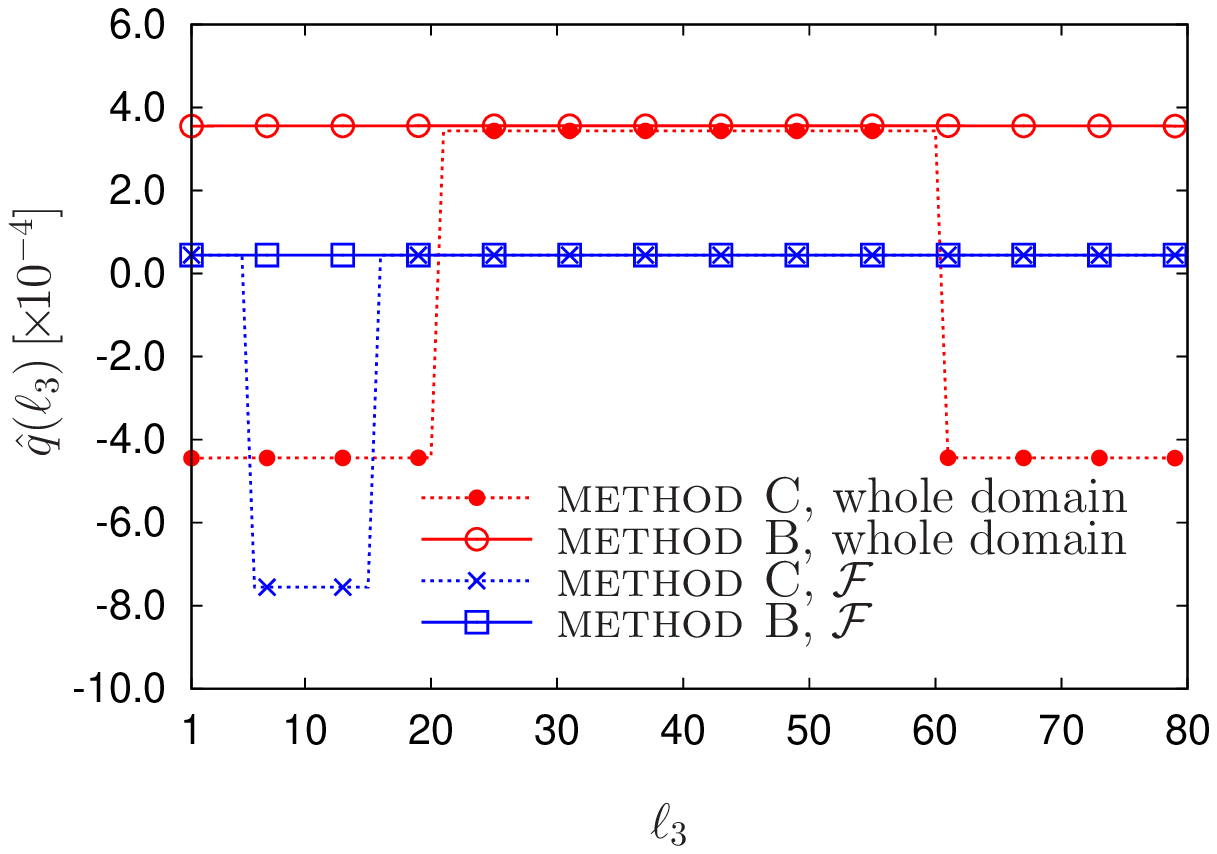}
   \end{centering}
  \caption{Mass flux $q(\lk)$ in a quadratic pipe of width $\hat{B}=5$ as
obtained from a \LBBGK\ simulation. If the
    acceleration is implemented as defined by \mtC, $q(\lk)$ is not
    constant in the regions where the acceleration is applied. \mtB\
    ensures a correct constant mass flux $q(\lk)$ throughout the whole
    domain.}
\label{plot_f}
\end{figure}

In Figure \ref{plot_f} we compare \mtB\ and \mtC.  
All simulation parameters except the pipe width
($\hat{B}=5$) are kept as before so that
\begin{equation}
\! \mc{S}\!=\!\{\bfl\!\in\!\mc{L}\! :\! 18\! \leq \li\! \leq 22 ,18\! \leq \lj \!\leq 22,  21\!\leq \lk \!\leq 60\}.\!
\end{equation}
The line representing the application of \mtC\ and the external acceleration
applied throughout the whole domain has discontinuities exactly at the
position where the width of the channel changes abruptly, i.e. at $\lk=20$ and $\lk=60$
(local porosity dependency). The line representing the application of \mtC\
throughout an acceleration zone $\mc{F}$ shows discontinuities within the
acceleration zone, i.e. in the interval $\lk\in[6,15]$. 
These discontinuities are not present when using \mtB.
The term $\eta \langle \vk \rangle_{s}$ in Darcy's law (see
Eq.~\eqref{darcy}) is usually approximated by $Q\nu/A$, where the total
mass flux $Q$ is calculated by averaging $q(\lk)$ in the whole
sample. $A$ represents the sample cross-sectional area.  If an
external acceleration is not implemented correctly, the calculated $Q$
is always incorrect leading to a wrong estimate of $\kappa$. For the
example in Fig.~\ref{plot_f} which uses \mtC\ and a force throughout
the whole domain, $Q$ is underestimated but remains positive. This is
not the case if an acceleration zone $\mc{F}$ is used. Here, the
calculation of $Q$ leads to an unphysical negative result, so that the
permeability is always underestimated and in some cases negative. Such
cases can also be observed for inhomogeneous stochastic porous media,
where the variation of pore sizes is very large~\cite{NRHH10}.

\begin{figure}[ht]
\includegraphics[width=0.9\columnwidth]{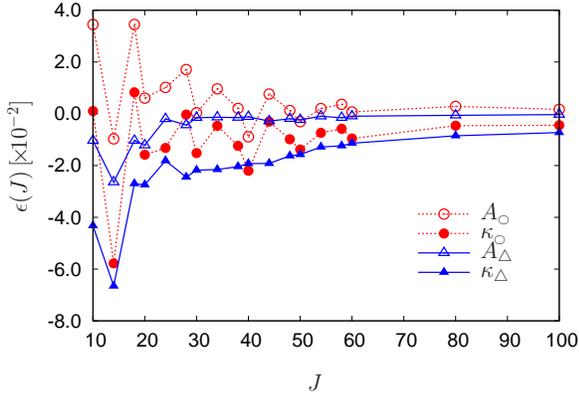}
\caption{Relative errors in the area discretization
  $\epsilon^{\sbgk}_{A}$ and permeability estimation
  $\epsilon^{\sbgk}_{\kappa}$ for a circular and triangular
  cross-section pipe with different system sizes $J$. The permeability
  error correlates with the discretization error. Compared to the
  results for quadratic pipes, the permeability error is at the same
  resolution approximately twice as large.  }
 \label{fig:trianglecircle} 
\end{figure}

Another important issue is the effect of discretization. 
When investigating square pipes the lattice is aligned with the
solid-fluid interface. This is not the case for the simulation of flow
in realistic stochastic porous media. Thus, the influence of
discretization effects is substantially larger than in the ideal cases
presented before. We investigate the order of the resulting error by
calculating the permeabilities in pipes with a circular and an
equilateral triangular cross-section.
The samples are of size $\Li=J$, $\Lj=J$ and, $\Lk=4J$ with $J\in
\mathbbm{N}$ and the cross-sections are defined by their diameter
$B_{\crcl}$ (circular) or their side-length $B_{\triangle}$
(equilateral triangle) with $B_{\crcl}=B_{\triangle}=(J-2)\dx$.  The
analytical solutions for the permeabilities of circular and
equilateral triangular pipes are
\begin{eqnarray}
\label{kcircle}
\kappa^{\thry}_{\crcl}&=&\frac{A^{\thry}_{\crcl}}{8\pi}, 
\quad  
A^{\thry}_{\crcl}=\frac{\pi}{4}\,{B_{\crcl}}^{2},\\
\label{ktri} 
\kappa^{\thry}_{\triangle}&=&\frac{\sqrt{3} A^{\thry}_{\triangle}}{60}, 
\quad A^{\thry}_{\triangle}=\frac{\sqrt{3}}{4}\,{B_{\triangle}}^{2},
\end{eqnarray}
where $A^{\thry}_{\crcl}$ and $A^{\thry}_{\triangle}$ are the
cross-sectional areas.  Discretizing these areas on a cubic lattice
results in approximate cross-sectional areas $A^{\sLB}_{\triangle}$ and
$A^{\sLB}_{\crcl}$.  Let $\epsilon^{\sLB}_{A_{\triangle}}$ and
$\epsilon^{\sLB}_{A_{\crcl}}$ be the relative discretization errors of
those areas. The permeabilities as calculated from the simulation
results are $\kappa^{\sbgk}_{\triangle}$ and $\kappa^{\sbgk}_{\crcl}$,
with their relative errors being
$\epsilon^{\sbgk}_{\kappa_{\triangle}}$ and
$\epsilon^{\sbgk}_{\kappa_{\crcl}}$.
Fig.~\ref{fig:trianglecircle} depicts that for both geometries the
relative error of the permeabilities is much larger than for pipes
with quadratic cross-section, see Fig.~\ref{fig:mrtvssrt2} for
comparison. Furthermore, it can be seen that the error in permeability
correlates with the error of the discretized area.  This discretization error is not present when investigating
square pipes that are aligned with the grid. In stochastic porous
media this discretization error is inevitable. Therefore, arbitrarily
structured pore throats have to be resolved at a much higher
resolution for high precision permeability calculations. This is a
serious limitation when calculating permeabilities for laboratory
sized porous media using the techniques discussed in this article.

\section{Application to Fontainebleau sandstones}
After validating the simulation results, determining errors for pipe
flow, and pointing out problems when calculating permeabilities with
\LB\ implementations, we now apply our findings to investigate a
porous sample.  We calculate the permeability of a sample of
Fontainebleau sandstone, gained by thresholding a discretized \muCT\
data set. This particular data is chosen, because it
has been investigated previously using a finite difference method and
another \LB-implementation~\cite{MAKHT02}. The results for the
calculated permeabilities in~\cite{MAKHT02} show excellent agreement
with the experimental results in~\cite{Man01}, therefore this sample
serves as a benchmark for the permeability calculations presented
here.
Calculations are carried out at a low (\lr) and high resolution
(\hr). The \lr\ computational domain is
\begin{equation}
\label{fntb_orig}
\begin{split}
\Li&=305, \, \Lj=305, \, \Lk=320,\\
\mathcal{S}&=\{\bfl\in\mc{L} : 3\leq \li\leq 303,  
3\leq \lj\leq 303,11\leq \lk\leq 310\},\\
\mathcal{I}&=\{\bfl\in\mc{L} : 1\leq \lk\leq 10\},
\mathcal{O}=\{\bfl\in\mc{L} : 311\leq \lk\leq 320\},\\
\mathcal{F}&=\{\bfl\in\mc{L} : 1\leq \lk\leq 5\},
\Lin=10,\,\Ls=300,\,\Lout=10,\\
\dx &= 7.5\pwr{-6}\,\mathrm{m}.\nonumber
\end{split}
\end{equation}
The high resolution sample is created from the low resolution sample by
substituting every voxel with eight voxels on a cubic sublattice. The
\hr\ computational domain is
\begin{equation}
\begin{split}
\Li&=605, \, \Lj=605, \, \Lk=620,\\
\mathcal{S}&=\{\bfl\in\mc{L} : 3\leq \li\leq 602 ,
3\leq \lj\leq 602, 11\leq \lk\leq 610\},\\
\mathcal{I}&=\{\bfl\in\mc{L} : 1\leq \lk\leq 10\},
\mathcal{O}=\{\bfl\in\mc{L} : 611\leq \lk\leq 620\},\\
\mathcal{F}&=\{\bfl\in\mc{L} : 1\leq \lk\leq 5\},
\Lin=10,\,\Ls=600,\,\Lout=10,\\
\dx &= 3.75\pwr{-6}\,\mathrm{m}.\nonumber
\end{split}
\end{equation}
For the permeability calculation an approximation of Darcy's law
is used, see Eqs.~\eqref{darcy},~\eqref{dpp} and~\eqref{rhop}, 
\begin{equation}
\label{sanddarcy} 
\kappa \! = \! - \frac{
\nu (\Ls\!-\!1) \dx  \, {\langle \vk \rangle}_{\mc{S}}
({\langle\rho\rangle}_{\Cinp \cap \mc{P}}\!+\!{\langle\rho\rangle}_{\Coutm \cap \mc{P}})
}
{  2 \, ( {\langle p \rangle}_{\Coutm \cap \mc{P}}\!-\!{\langle p \rangle}_{\Cinp \cap \mc{P}} )  }.\!
\end{equation}
The kinematic viscosity $\nu$ is calculated using Eq.~\eqref{visc_BGK}
with $\htau$, as in Tab.~\ref{tbl:fntb}, $\htaub=1.0$ and $\dt$ from
Eq.~\eqref{cs} with $\cs=1 \mathrm{m}\,\mathrm{s}^{-1}$. Simulations
are performed with the \LBBGK\ and \LBMRT\ method for $100\,000$
time steps.
The relaxation times used in the simulation and the calculated
permeabilities for the \lr, \hr\ sample and
from~\cite{MAKHT02} are given in Tab.~\ref{tbl:fntb}.  The calculated
permeabilities $\kappa_{\lr}$ and $\kappa_{\hr}$
were linearly extrapolated for infinite resolution at $1/ \Ls=0$
yielding $\kappa_{\extrap}$.

\begin{figure}[ht]
\begin{centering}
\includegraphics[width=0.9\columnwidth]{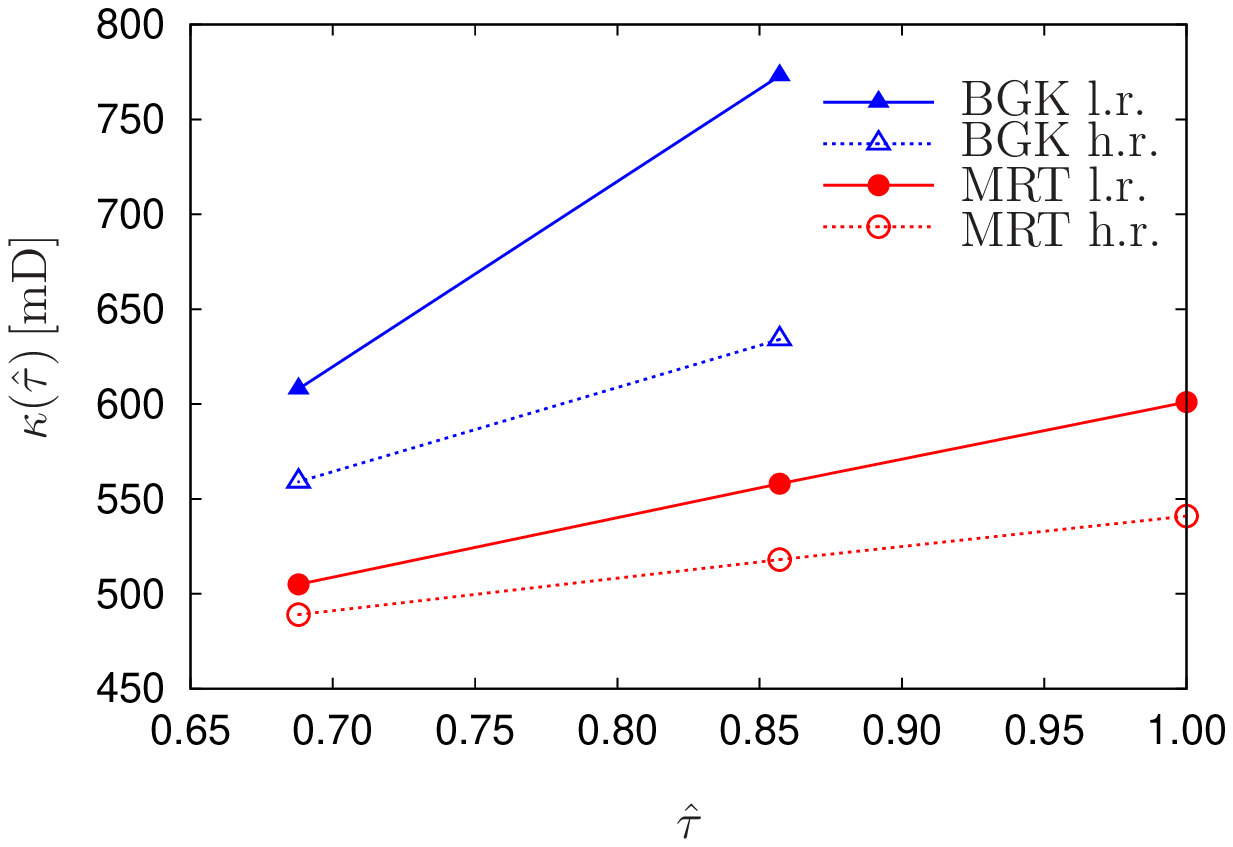}
\includegraphics[width=0.9\columnwidth]{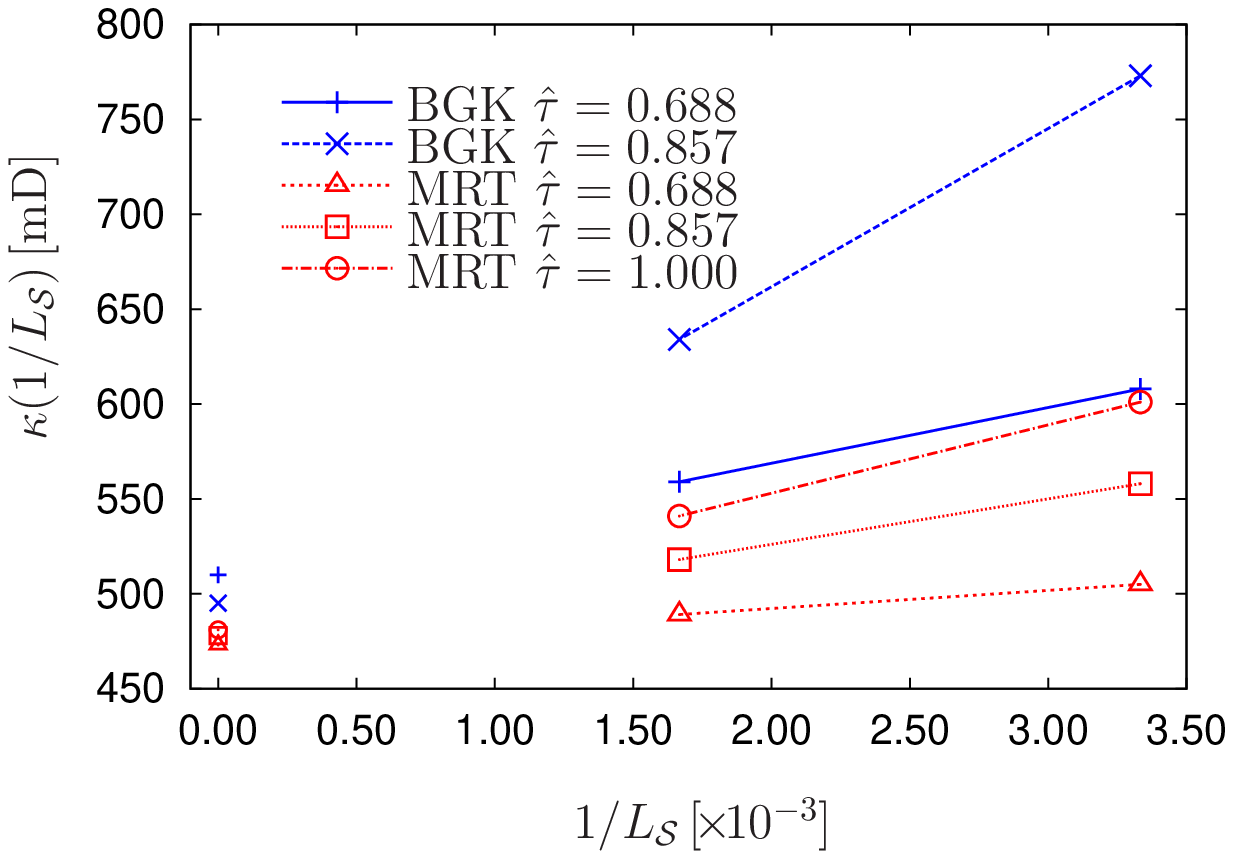}
\end{centering}
\caption{\label{fig:fntb} The top plot shows the permeability for different
$\htau$, \LB\ schemes, the \lr\ and the \hr\ sample. The
influence of $\htau$ on the permeability is stronger using the \BGK\
implementation.  The bottom plot shows the permeability results, for the
\lr\ sample with $\Ls=300$ and the \hr\ sample with
$\Ls=600$. At $1/ \Ls=0$ the extrapolated permeabilities
$\kappa_{\extrap}$ are shown.}
\end{figure}

Our result $\kappa_{\lr}=608 \, [\mD]$ for $\htau=0.688$,
see Tab.~\ref{tbl:fntb}, is in good agreement with the result
in~\cite{MAKHT02} being $\kappa=621\,[\mD]$. Although the simulation
setup is different, a relative difference of only $2\%$ is obtained.
Fig.~\ref{fig:fntb} (top) confirms that permeability results gained
from \LBBGK\ simulations are particularly dependent on the relaxation
time $\htau $ that is used. Therefore, when
investigating complex geometries, where $\htau$ cannot be optimized for
a specific geometrical shape, a \LBMRT\ should be used. As expected,
when calculating permeabilities for complex geometries, the influence
of $\htau$ is much stronger than within simple geometries, i.e.,
square pipes, see Secs.~\ref{lcalib:sec} and~\ref{diffperm:sec}
together with Fig.~\ref{permtau_bgkmrt_fig}.
The extrapolated permeabilities $\kappa_{\extrap}$ are an
estimate for the true permeability of the discretized \muCT\ sample
at resolution $3.75\pwr{-6}\,\mathrm{m}$ and not the true permeability
of the sandstone. To estimate the true permeability of the sandstone
by extrapolation, new \muCT\ data with higher resolutions would be
required. However, from Fig.~\ref{fig:fntb} (bottom) it can be seen
that the extrapolated permeability values have a small spread, in a
range from 473--$510\,[\mD]$ regardless of the simulation method and
relaxation time used. This indicates that, if sufficiently high
resolved sample data and computer performance are available, an
extrapolation analysis, even using \LBBGK\ $\htau=1$ results,
might give a good approximation of the true permeability. 
{ 
One could expect that errors due to
random geometries would experience random cancellations. However, the
results presented in Fig.~\ref{fig:fntb} clearly show that this is not true for a
realistic porous medium.
}
\begin{table}[htb]
\begin{centering}
\begin{tabular}{|c|c|c|c|c|} 
\hline \LB\ Method & $\htau$ & $\kappa_{\lr}\,[\mD]$ &
$\kappa_{\hr}\,[\mD]$ &
$\kappa_{\extrap}\,[\mD]$\tabularnewline \hline \hline
\BGK~\cite{MAKHT02}& 0.688 & 621 & --- & --- \tabularnewline \hline
\BGK & 0.688 & 608 & 559 & 510 \tabularnewline \hline \BGK & 0.857 &
773 & 634 & 495 \tabularnewline \hline \MRT & 0.688 & 505 & 489 & 473
\tabularnewline \hline \MRT & 0.857 & 558 & 518 & 478 \tabularnewline
\hline \MRT & 1.000 & 601 & 541 & 481 \tabularnewline \hline
\end{tabular}
\end{centering}
\caption{\label{tbl:fntb} Simulation results of the permeability
  calculations for the Fontainebleau sandstone.  Results for the low resolution
sample are labeled $\kappa_{\lr}$, the high resolution sample results
$\kappa_{\hr}$ and the extrapolation results
$\kappa_{\extrap}$.  Different relaxation-times $\htau$ and
\LBBGK\ and \LBMRT\ implementations are compared.}
\end{table}

\section{Conclusion} 
Our simulation setup of an acceleration zone and in/outlet
chambers, together with our approximations of Darcy's law, provides a
 method for permeability calculations.
Several problems in the numerical implementation and data evaluation
were addressed, such as a correct acceleration implementation and an
adequate approximation for calculating the pressure gradient. Caveats
when using \LB\ simulations to calculate permeabilities have been
exposed.
We performed detailed studies with different \LB-implementations,
i.e., \BGK\ and \MRT, and for various systems to quantitatively
determine the accuracy of the calculated velocity field and calculated
permeability. We find that for reasonably resolved quadratic pipes, the
error of the calculated permeability is below $1\%$. Investigating
non-aligned geometries, circular and triangular pipes, the
discretization and permeability error is roughly $4\%$ at comparable
resolutions. From this we infer that permeability calculations in
stochastic porous media will have a significantly larger error, because the
resolution of pores and pore walls is usually well below the
resolution used for our pipe calculations above.
Comparing the two \LB-implementations, \LBBGK\ and \LBMRT, we find that
\LBMRT\ reduces the dependence of the permeability on the value of $\tau$
substantially. Using \LBBGK\ and a relaxation time $\tau$ 
tailored to give good results for a specific geometry does not assure reliable
results in a stochastic porous medium. For example, we found that 
\LBBGK\ and $\hat{\tau}=0.857$ yields the best result for 3D Poiseuille
flow in a quadratic pipe. However, for this value \LBBGK\ and \LBMRT\ results
differ by 20\% if applied to the Fontainebleau sandstone (see
Fig.~\ref{fig:fntb}). 
Therefore, \LBMRT\ is suggested to be used for
permeability estimates based on \LB\ simulations.
Further investigations using the \LB\ method for flow through
stochastic porous media should include a resolution and relaxation
time dependent analysis together with an appropriate extrapolation
scheme for more reliable permeability estimates.

\section{Acknowledgments}
We are grateful to the High Performance Computing Center in Stuttgart,
the Scientific Supercomputing Center in Karlsruhe, and the J\"{u}lich
Supercomputing Center for providing access to their machines.  
One of us (T.Z.) would like to acknowledge partial support 
from the DFG program EXC310 (Simulationstechnik).
We would like to thank Bibhu Biswal for fruitful discussions and the 
Sonderforschungsbereich~716, the DFG program ``Nano- and microfluidics'', 
and the Deutscher Akademischer Austauschdienst (DAAD) for financial support.


\end{document}